\documentclass[review,10pt,3p]{elsarticle}

\usepackage{amssymb,amsmath}
\usepackage{subfigure,tabularx}
\usepackage{txfonts}
\usepackage{color,ulem}
\usepackage[colorlinks,hyperindex]{hyperref}

\newcommand{\figdraft}{false} 

\newcommand{\eq}{\text{eq}}
\newcommand{\T}{\text{T}}
\newcommand{\hu}{\hat{u}}
\newcommand{\hbu}{\hat{\mathbf{u}}}
\newcommand{\hF}{\hat{F}}
\newcommand{\hbF}{\hat{\mathbf{F}}}
\newcommand{\tG}{\tilde{G}}
\newcommand{\tbG}{\tilde{\mathbf{G}}}
\newcommand{\rhor}{\rho_0^{}}
\newcommand{\bbm}{\bar{\mathbf{m}}}
\newcommand{\LBE}{\text{\tiny LBE}}
\newcommand{\EOS}{\text{\tiny EOS}}
\newcommand{\Ma}{M\mspace{-1.5mu}a}
\newcommand{\Prt}{P\mspace{-1.5mu}r}
\newcommand{\Ec}{E\mspace{-1.5mu}c}
\newcommand{\Ra}{R\mspace{-1.5mu}a}
\newcommand{\Nu}{N\mspace{-1.5mu}u}
\newcommand{\ff}{{f\mspace{-4.0mu}f}}

\begin{document}

\begin{frontmatter}

\title{Lattice Boltzmann model with self-tuning equation of state for
coupled thermo-hydrodynamic flows}

\author[SJTU,TUM]{Rongzong Huang}
\ead{rongzong.huang@tum.de}
\author[SJTU]{Huiying Wu}
\ead{whysrj@sjtu.edu.cn}
\author[TUM]{Nikolaus A. Adams}
\ead{nikolaus.adams@tum.de}

\address[SJTU]{School of Mechanical Engineering, Shanghai Jiao
Tong University, 200240 Shanghai, China}
\address[TUM]{Institute of Aerodynamics and Fluid Mechanics, Technical
University of Munich, 85748 Garching, Germany}

\begin{abstract}
A novel lattice Boltzmann (LB) model with self-tuning equation of state
(EOS) is developed in this work for simulating coupled thermo-hydrodynamic
flows. The velocity field is solved by the recently developed
multiple-relaxation-time (MRT) LB equation for density distribution
function (DF), by which a self-tuning EOS can be recovered. As to the
temperature field, a novel MRT LB equation for total energy DF is directly
developed at the discrete level. By introducing a density-DF-related term
into this LB equation and devising the equilibrium moment function for
total energy DF, the viscous dissipation and compression work are
consistently considered, and by modifying the collision matrix, the
targeted energy conservation equation is recovered without deviation term.
The full coupling of thermo-hydrodynamic effects is achieved via the
self-tuning EOS and the viscous dissipation and compression work. The
present LB model is developed on the basis of the standard lattice, and
various EOSs can be adopted in real applications. Moreover, both the
Prandtl number and specific heat ratio can be arbitrarily adjusted.
Furthermore, boundary condition treatment is also proposed on the basis of
the judicious decomposition of DF into its equilibrium, force (source),
and nonequilibrium parts. The local conservation of mass, momentum, and
energy can be strictly satisfied at the boundary node. Numerical
simulations of thermal Poiseuille and Couette flows are carried out with
three different EOSs, and the numerical results are in good agreement with
the analytical solutions. Then, natural convection in a square cavity with
a large temperature difference is simulated for the Rayleigh number from
$10^3$ up to $10^8$. Good agreement between the present and previous
numerical results is observed, which further validates the present LB
model for coupled thermo-hydrodynamic flows.
\end{abstract}


\end{frontmatter}

\section{Introduction} \label{sec.introduction}
The lattice Boltzmann (LB) method has developed into an attractive numerical
method over the past three decades for simulating complex fluid flows
\cite{Aidun2010, Gross2010, Huang2016.amr} and solving various partial
differential equations \cite{Ginzburg2005, Dellar2011, Chai2018}.
Historically, the LB method originates from the lattice gas automata (LGA)
to eliminate the statistical noise \cite{McNamara1988}, and thus it inherits
some distinguishing features from LGA, such as the simple algorithm (local
collision and linear streaming) and the easy incorporation of microscopic
interactions \cite{Shan1993, Ladd1994}. Afterward, it is found that the
classical LB model for hydrodynamic flows can be derived from the
Boltzmann-BGK equation via systematic discretization \cite{He1997a,
He1997b}, and then various LB models for multiphase flows
\cite{He1998.nonideal, Luo1998} and thermo-hydrodynamic flows (i.e., thermal
fluid flows) \cite{He1998.thermal, Guo2007} have been established from the
kinetic models in an {\it a priori} manner.
\par

Since most hydrodynamic flows involve some forms of thermal effects,
thermo-hydrodynamic flows are extensively encountered in nature and
engineering, and the LB method for simulating thermo-hydrodynamic flows has
attracted continuous attention since the early 1990s \cite{He1998.thermal,
Guo2007, Shan1997, Alexander1993, McNamara1997, Zheng2008, Lallemand2003,
Mezrhab2004, Mezrhab2010, Huang2014.ibm, Contrino2014}. However, it remains
open-ended though the LB method has achieved great success in simulating
isothermal fluid flows \cite{Sbragaglia2009, Succi2015}. Generally, the
existing LB models for thermo-hydrodynamic flows can be categorized into
three major groups: the multispeed approach \cite{McNamara1997, Zheng2008},
the double-distribution-function (DDF) approach \cite{He1998.thermal,
Guo2007, Shan1997}, and the hybrid approach \cite{Lallemand2003,
Mezrhab2004}. The multispeed approach uses a single distribution function
(DF) to describe the mass, momentum, and energy conservation laws, and thus
it requires more discrete velocities than the standard lattice (i.e., it
requires the multispeed lattice). By definition, the DDF approach consists
of double DFs, with one DF for the mass and momentum conservation laws and
the other DF for the energy conservation law. In the hybrid approach, the
mass and momentum conservation laws are described by one DF, while the
energy conservation law is described by a macroscopic governing equation
that is solved via the conventional computational fluid dynamics methods.
Severe numerical instability \cite{Lallemand2003} and complexity of boundary
condition treatment \cite{Lee2018} are usually encountered in the multispeed
approach. As to the hybrid approach, it acts as a compromised solution that
deviates from the mesoscopic LB method \cite{Lallemand2003}, and the viscous
dissipation is usually ignored in this approach \cite{Feng2018, Safari2018}.
On the contrary, the DDF approach, free of the above drawbacks, is most
widely studied and adopted in real applications.
\par

Most of the existing DDF LB models for thermo-hydrodynamic flows are
inherently a decoupling model, which means that the recovered equation of
state (EOS) is a decoupling EOS $p_\EOS^{} = \rho R_g T_0$ ($R_g$ is the gas
constant and $T_0$ is the reference temperature), where the pressure is not
directly related to the temperature. Consequently, these LB models are
restricted to the thermo-hydrodynamic flows under the Boussinesq
approximation (i.e., the decoupling thermo-hydrodynamic flows). Based on the
DDF kinetic model constructed by Guo et al.\ \cite{Guo2007}, and by applying
the discretization of velocity space presented by Shan et al.\
\cite{Shan2006} that can lead to the temperature-independent discrete
velocities, Hung and Yang \cite{Hung2011} proposed a DDF LB model aimed at
recovering the ideal-gas EOS $p_\EOS^{} = \rho R_g T$. However, the
deviation in the third-order moment of the equilibrium distribution function
(EDF) for density due to the constraint of standard lattice, as previously
identified by Prasianakis and Karlin \cite{Prasianakis2007}, is not
considered in Hung and Yang's model, and meanwhile, an error also exists in
their derived EDF for total energy. In 2012, by introducing the correction
term for the third-order moment of the EDF for density and deriving the
correct EDF for total energy, Li et al.\ \cite{Li2012} developed a DDF LB
model for simulating coupled thermo-hydrodynamic flows. The ideal-gas EOS
can be recovered by Li et al.'s model, and the simulation of natural
convection with a large temperature difference is reported \cite{Li2012}.
Following the similar way, Feng et al.\ \cite{Feng2015} proposed
three-dimensional DDF LB models. A correction term for the second-order
moment of the EDF for total energy is further introduced by Feng et al.\
\cite{Feng2015} to enhance the numerical stability of the LB equation for
total energy DF. Recently, the cascaded collision scheme is employed in the
LB equation for density DF to enhance the numerical stability by Fei and Luo
\cite{Fei2018}, while the single-relaxation-time (SRT) collision scheme is
still used in the LB equation for total energy DF.
\par

It is worth pointing out that the ideal-gas EOS is recovered by the above
DDF LB models \cite{Hung2011, Li2012, Feng2015, Fei2018}, which indicates
that these models are only applicable to the coupled thermo-hydrodynamic
flows of ideal gases. Moreover, in these models, the LB equation for total
energy DF is complicated due to the consideration of the viscous dissipation
and compression work, and thus it is difficult to employ the
multiple-relaxation-time (MRT) or cascaded collision schemes in this LB
equation to enhance the numerical stability although the MRT and cascaded
collision schemes have been employed in the LB equation for density DF
\cite{Li2012, Fei2018}. Most recently, we developed an LB model with
self-tuning EOS for multiphase flows \cite{Huang2018.eos}. Since the
recovered EOS can be self-tuned via a built-in variable, this model serves
as a good and distinct starting point for developing a novel LB model for
coupled thermo-hydrodynamic flows, which is the main objective of the
present work. To be specific, a novel MRT LB equation for solving the energy
conservation equation, with considering the viscous dissipation and
compression work, is developed. Furthermore, boundary condition treatment
for simulating coupled thermo-hydrodynamic flows is also proposed on the
basis of the judicious decomposition of DF into three parts rather than two.
The remainder of the present paper is organized as follows. In Section
\ref{sec.lbm}, a novel LB model for coupled thermo-hydrodynamic flows is
developed. In Section \ref{sec.bc}, boundary condition treatment is
proposed. Numerical validations of the present LB model are carried out in
Section \ref{sec.validation}, and a brief conclusion is drawn in Section
\ref{sec.conclusion}.
\par

\section{Lattice Boltzmann model} \label{sec.lbm}
The present LB model for coupled thermo-hydrodynamic flows is developed on
the basis of the recent LB model with self-tuning EOS for multiphase flows.
Double DFs are involved: one is the density DF used to solve the velocity
field (i.e., the mass-momentum conservation equations), and the other is the
total energy DF used to solve the temperature field (i.e., the energy
conservation equation). The full coupling of thermo-hydrodynamic effects is
achieved via the self-tuning EOS recovered by the LB equation for density DF
and the viscous dissipation and compression work considered in the LB
equation for total energy DF. Both the LB equations for density and total
energy DFs are based on the standard lattice. For the sake of simplicity and
clarity, the two-dimensional model will be developed here, and its extension
to three-dimensional model is straightforward. The standard two-dimensional
nine-velocity (D2Q9) lattice is given as \cite{Qian1992}
\begin{equation}
\mathbf{e}_i =
\begin{cases}
c \big( 0 ,\, 0 \big) ^\T, & i=0, \\
c \big( \cos[(i-1)\pi/2] ,\, \sin[(i-1)\pi/2] \big) ^\T,
                           & i=1,2,3,4, \\
\sqrt{2}c \big( \cos[(2i-1)\pi/4] ,\, \sin[(2i-1)\pi/4] \big) ^\T,
                           & i=5,6,7,8, \\
\end{cases}
\end{equation}
where the lattice speed $c = \delta_x / \delta_t$ with $\delta_x$ and
$\delta_t$ being the lattice spacing and time step, respectively.
\par

\subsection{LB equation for density DF}
The recently developed LB equation for density DF that can recover a
self-tuning EOS is briefly introduced here for self-completeness. The MRT LB
equation for density DF $f_i (\mathbf{x}, t)$ can be expressed as
\cite{Huang2018.eos}
\begin{subequations}\label{eq.LBE.u}
\begin{equation}\label{eq.LBE.u.s}
f_i (\mathbf{x} + \mathbf{e}_i \delta_t, t + \delta_t)
= \bar{f} _i (\mathbf{x}, t),
\end{equation}
\begin{equation}\label{eq.LBE.u.c}
\bar{ \mathbf{m} } (\mathbf{x}, t) = \mathbf{m} + \delta_t \mathbf{F}_m -
\mathbf{S} \left( \mathbf{m} - \mathbf{m}^\eq + \dfrac{\delta_t}{2}
\mathbf{F}_m \right) - \mathbf{R} \left( \mathbf{I} -
\dfrac{\mathbf{S}}{2}
\right) \left( \mathbf{m} - \mathbf{m}^\eq + \dfrac{\delta_t}{2}
\mathbf{F}_m \right) - \delta_x \mathbf{T} \cdot \nabla \rho -
\dfrac{\delta_x}{c^2} \mathbf{X} \cdot \nabla p_\LBE^{},
\end{equation}
\end{subequations}
where Eq.\ (\ref{eq.LBE.u.s}) is the streaming process executed in velocity
space and Eq.\ (\ref{eq.LBE.u.c}) is the collision process executed in
moment space at position $\mathbf{x}$ and time $t$. The moment of density DF
in Eq.\ (\ref{eq.LBE.u.c}) is given as $\mathbf{m} = \mathbf{M} (f_i)^\T$.
Here, $\mathbf{M}$ is the dimensionless transformation matrix
\cite{Lallemand2000}
\begin{equation}\label{eq.M}
\mathbf{M} =
\begin{bmatrix}
 1& 1& 1& 1& 1& 1& 1& 1& 1 \\
-4&-1&-1&-1&-1& 2& 2& 2& 2 \\
 4&-2&-2&-2&-2& 1& 1& 1& 1 \\
 0& 1& 0&-1& 0& 1&-1&-1& 1 \\
 0&-2& 0& 2& 0& 1&-1&-1& 1 \\
 0& 0& 1& 0&-1& 1& 1&-1&-1 \\
 0& 0&-2& 0& 2& 1& 1&-1&-1 \\
 0& 1&-1& 1&-1& 0& 0& 0& 0 \\
 0& 0& 0& 0& 0& 1&-1& 1&-1 \\
\end{bmatrix},
\end{equation}
and $(f_i)^\T$ denotes the vector $(f_0, f_1, \cdots, f_8)^\T$. The
post-collision density DF in Eq.\ (\ref{eq.LBE.u.s}) is obtained via the
inverse transformation $( \bar{f}_i )^\T = \mathbf{M}^{-1} \bbm$, and the
post-collision moment $\bbm$ is computed by Eq.\ (\ref{eq.LBE.u.c}). The
last three terms on the right-hand side (RHS) of Eq.\ (\ref{eq.LBE.u.c}) are
the correction terms aimed at eliminating the additional cubic terms of
velocity in the recovered momentum conservation equation
\cite{Huang2018.cubic}, where $p_\LBE^{}$ denotes the recovered EOS by the
LB equation. The macroscopic density $\rho$ and velocity $\mathbf{u}$ are
defined as
\begin{equation}\label{eq.rho.u}
\rho = \sum\nolimits_i f_i, \quad
\rho \mathbf{u} = \sum\nolimits_i \mathbf{e}_i f_i + \dfrac{\delta_t}{2}
\mathbf{F},
\end{equation}
where $\mathbf{F}$ is the force term. In the recent LB model for multiphase
flows \cite{Huang2018.eos}, $\mathbf{F}$ is the total force due to the
long-range molecular interaction, while in the present LB model for coupled
thermo-hydrodynamic flows, $\mathbf{F}$ is simply an external force, such as
the gravity force.
\par

In Eq.\ (\ref{eq.LBE.u.c}), the equilibrium moment function for density DF
$\mathbf{m} ^\eq$ is given as \cite{Huang2018.eos}
\begin{equation}
\begin{split}
\mathbf{m}^\eq = \big[ \rho ,\,
2\alpha_1^{} \rho + 2\beta_1^{} \eta + 3 \rho |\hbu|^2 ,\,
\alpha_2^{} \rho + \beta_2^{} \eta - 3 \rho |\hbu|^2 + 9 \rho \hu_x^2
\hu_y^2 &,\,
\\
\rho \hu_x ,\,
-\rho \hu_x + 3 \rho \hu_x \hu_y^2 ,\,
\rho \hu_y ,\,
-\rho \hu_y + 3 \rho \hu_y \hu_x^2 &,\,
\rho (\hu_x^2 - \hu_y^2) ,\,
 \rho \hu_x \hu_y \big] ^\T ,
\end{split}
\end{equation}
where $\hbu = \mathbf{u} /c$ and $\eta$ is the built-in variable aimed at
achieving a self-tuning EOS. The coefficients $\alpha_1^{}$ and $\beta_1^{}$
are set to $-1$ and $1$, respectively, while the coefficients $\alpha_2^{}$
and $\beta_2^{}$ are determined by Eq.\ (\ref{eq.LBE.u.coefficients}). The
discrete force term in moment space $\mathbf{F}_m$ is given as
\begin{equation}
\begin{split}
\mathbf{F}_m = \big[ 0 ,\,
6 \hbF \cdot \hbu ,\,
-6 \hbF \cdot \hbu + 9[\hbF \hbu \hbu \hbu]_{xxyy} ,\,
\hF_x ,\,
-\hF_x + 3[\hbF \hbu \hbu]_{xyy} &,\,
\\
\hF_y ,\,
-\hF_y + 3[\hbF \hbu \hbu]_{xxy} &,\,
2 (\hF_x \hu_x - \hF_y \hu_y) ,\,
\hF_x \hu_y + \hF_y \hu_x \big] ^\T ,
\end{split}
\end{equation}
where $\hbF = \mathbf{F} /c$, and the square bracket and its subscript
denote permutation and tensor index, respectively. For example, $[\hbF \hbu
\hbu \hbu]_{xxyy} = 2 \hF_x \hu_x \hu_y^2 + 2 \hF_y \hu_y \hu_x^2$. To
correctly recover the Newtonian viscous stress tensor, the collision matrix
in moment space $\mathbf{S}$ is modified as follows \cite{Huang2018.eos}
\begin{equation}
\mathbf{S} =
\begin{bmatrix}
s_0^{}& 0& 0& 0& 0& 0& 0& 0& 0\\
0& s_e^{}& k s_\varepsilon^{} \omega_e^{}&
            0& h \hu_x s_q^{} \omega_e^{}&
            0& h \hu_y s_q^{} \omega_e^{}& 0& 0\\
0& 0& s_\varepsilon^{}& 0& 0& 0& 0& 0& 0\\
0& 0& 0& s_j^{}& 0& 0& 0& 0& 0\\
0& 0& 0& 0& s_q^{}& 0& 0& 0& 0\\
0& 0& 0& 0& 0& s_j^{}& 0& 0& 0\\
0& 0& 0& 0& 0& 0& s_q^{}& 0& 0\\
0& 0& 0& 0&  2 b \hu_x s_q^{} \omega_p^{}&
         0& -2 b \hu_y s_q^{} \omega_p^{}& s_p^{}& 0\\
0& 0& 0& 0&    b \hu_y s_q^{} \omega_p^{}&
         0&    b \hu_x s_q^{} \omega_p^{}& 0& s_p^{}\\
\end{bmatrix},
\end{equation}
where $\omega_{e, p}^{} = s_{e, p}^{} /2 -1$, and $k$, $h$, and $b$ are the
coefficients. Through the Chapman-Enskog analysis, the coefficients in
$\mathbf{m}^\eq$ and $\mathbf{S}$ should satisfy the following relations
\begin{equation}\label{eq.LBE.u.coefficients}
\alpha_2^{} = - \dfrac{2\alpha_1^{} + \varpi +1}{1-\varpi} ,\quad
\beta_2^{} = - \dfrac{2 \beta_1^{}}{1-\varpi} ,\quad
k = 1-\varpi ,\quad
h = \dfrac{6 \varpi (1-\varpi)}{1-3\varpi} ,\quad
b = \dfrac{1-\varpi}{1-3\varpi} ,
\end{equation}
where $\varpi$ is related to the bulk viscosity.
\par

In Eq.\ (\ref{eq.LBE.u.c}), the last three terms, together with the
high-order terms of velocity in $\mathbf{m}^\eq$ and $\mathbf{F}_m$, are
introduced to eliminate the additional cubic terms of velocity
\cite{Huang2018.cubic}, which are not considered in the previous DDF LB
models for coupled thermo-hydrodynamic flows. The correction matrix
$\mathbf{R}$ is a $9 \times 9$ matrix and it is set as \cite{Huang2018.eos}
\begin{subequations}
\begin{equation}
\mathbf{R} =
\begin{bmatrix}
0& 0&      0& 0& 0& 0& 0& 0&      0\\
0& R_{11}& 0& 0& 0& 0& 0& R_{17}& R_{18}\\
0& 0&      0& 0& 0& 0& 0& 0&      0\\
0& 0&      0& 0& 0& 0& 0& 0&      0\\
0& 0&      0& 0& 0& 0& 0& 0&      0\\
0& 0&      0& 0& 0& 0& 0& 0&      0\\
0& 0&      0& 0& 0& 0& 0& 0&      0\\
0& R_{71}& 0& 0& 0& 0& 0& R_{77}& R_{78}\\
0& R_{81}& 0& 0& 0& 0& 0& R_{87}& R_{88}\\
\end{bmatrix},
\end{equation}
where the nonzero elements can be determined via the Chapman-Enskog analysis
as follows
\begin{equation}
\setlength{\arraycolsep}{0.6em}
\begin{array}{lll}
R_{11} = -\tfrac{ (9-15k-2h) s_e^{} }{4\varpi}  (\hu_x^2 + \hu_y^2), &
R_{17} = -\tfrac{ 3(9-3k+2h) s_p^{} (2-s_e^{}) }{4(2-s_p^{})}
                                                (\hu_x^2 - \hu_y^2), &
R_{18} = \tfrac{ 12(3k+h) s_p^{} (2-s_e^{}) }{2-s_p^{}} \hu_x \hu_y, \\
R_{71} = -\tfrac{ (3-4b) s_e^{} (2-s_p^{}) }{4\varpi (2-s_e^{})}
                                                (\hu_x^2 - \hu_y^2), &
R_{77} = -\tfrac{ 3(3+4b) s_p^{} }{4} (\hu_x^2 + \hu_y^2), &
R_{78} = 0, \\
R_{81} = \tfrac{ b s_e^{} (2-s_p^{}) }{\varpi (2-s_e^{})} \hu_x \hu_y, &
R_{87} = 0, &
R_{88} = 6 b s_p^{} (\hu_x^2 + \hu_y^2). \\
\end{array}
\end{equation}
\end{subequations}
The correction matrix $\mathbf{T}$ is set as \cite{Huang2018.eos}
\begin{subequations}
\begin{equation}
\mathbf{T} = \big(
\mathbf{0}   ,\;
\mathbf{T}_1 ,\;
\mathbf{0}   ,\;
\mathbf{0}   ,\;
\mathbf{0}   ,\;
\mathbf{0}   ,\;
\mathbf{0}   ,\;
\mathbf{T}_7 ,\;
\mathbf{T}_8 \big) ^\T,
\end{equation}
whose element is a vector implying that the dimensions of $\mathbf{T}$ are
$9 \times 2$. The nonzero elements in $\mathbf{T}$ can also be determined
via the Chapman-Enskog analysis as follows
\begin{equation}
 \mathbf{T}_1 = \tfrac{ 3(2-s_e^{}) }{2}
\begin{bmatrix}
(1-k) \hu_x^3 - (2k+h) \hu_x \hu_y^2 \\
(1-k) \hu_y^3 - (2k+h) \hu_x^2 \hu_y
\end{bmatrix} ,\quad
\mathbf{T}_7 = \tfrac{ 2-s_p^{} }{2}
\begin{bmatrix}
 \hu_x^3 + 2b \hu_x \hu_y^2 \\
-\hu_y^3 - 2b \hu_x^2 \hu_y
\end{bmatrix} , \quad
\mathbf{T}_8 = - \tfrac{ b(2-s_p^{}) }{2}
\begin{bmatrix}
 \hu_y^3 + 2 \hu_x^2 \hu_y \\
 \hu_x^3 + 2 \hu_x \hu_y^2
\end{bmatrix} .
\end{equation}
\end{subequations}
Similarly to $\mathbf{T}$, the correction matrix $\mathbf{X}$ is a $9 \times
2$ matrix and it is set as \cite{Huang2018.eos}
\begin{subequations}
\begin{equation}
\mathbf{X} = \big(
\mathbf{0}   ,\;
\mathbf{X}_1 ,\;
\mathbf{0}   ,\;
\mathbf{0}   ,\;
\mathbf{0}   ,\;
\mathbf{0}   ,\;
\mathbf{0}   ,\;
\mathbf{X}_7 ,\;
\mathbf{X}_8 \big) ^\T,
\end{equation}
where the nonzero elements are given as
\begin{equation}
\mathbf{X}_1 = \tfrac{ 9 (2k+h) (2-s_e^{}) }{2}
\begin{bmatrix}
\hu_x \hu_y^2 \\
\hu_x^2 \hu_y
\end{bmatrix} ,\quad
\mathbf{X}_7 = -3 b (2-s_p^{})
\begin{bmatrix}
 \hu_x \hu_y^2 \\
-\hu_x^2 \hu_y
\end{bmatrix} ,\quad
\mathbf{X}_8 = \tfrac{ 3 b (2-s_p^{})}{2}
\begin{bmatrix}
\hu_y^3 + 2 \hu_x^2 \hu_y \\
\hu_x^3 + 2 \hu_x \hu_y^2
\end{bmatrix} .
\end{equation}
\end{subequations}
Here, we would like to point out that for the coupled thermo-hydrodynamic
flows under the low Mach number condition, these correction terms for the
additional cubic terms of velocity can be simply ignored. However, they are
kept in the present work for the sake of theoretical completeness and
computational accuracy.
\par

Through the Chapman-Enskog analysis, the following mass-momentum
conservation equations can be recovered \cite{Huang2018.eos}
\begin{equation}
\begin{cases}
\partial_t \rho + \nabla \cdot (\rho \mathbf{u}) = 0, \\
\partial_t (\rho \mathbf{u}) + \nabla \cdot (\rho \mathbf{uu}) =
-\nabla p_\LBE^{} + \mathbf{F} + \nabla \cdot \mathbf{\Pi}, \\
\end{cases}
\end{equation}
where $p_\LBE^{}$ and $\mathbf{\Pi}$ are the recovered EOS and viscous
stress tensor
\begin{equation}\label{eq.NSE}
p_\LBE^{} = c_s^2 [ (2+\alpha_1^{}) \rho + \beta_1^{} \eta ] ,\quad
\mathbf{\Pi} = \rho \nu [\nabla \mathbf{u} +
\mathbf{u} \nabla - (\nabla \cdot \mathbf{u}) \mathbf{I}] + \rho
\varsigma
(\nabla \cdot \mathbf{u}) \mathbf{I} ,
\end{equation}
where the lattice sound speed $c_s = c / \sqrt{3}$, the kinematic viscosity
$\nu = c_s^2 \delta_t \big( s_p^{-1} - 0.5 \big)$, and the bulk viscosity
$\varsigma = \varpi c_s^2 \delta_t \big( s_e^{-1} - 0.5 \big)$. As seen in
Eq.\ (\ref{eq.NSE}), the recovered EOS $p_\LBE^{}$ can be arbitrarily tuned
via the built-in variable $\eta$.
\par

\subsection{LB equation for total energy DF}
Since the EOS recovered by the above LB equation for solving the velocity
field can be self-tuned, we are now well equipped to simulate coupled
thermo-hydrodynamic flows. The remaining task is to develop an LB equation
for solving the temperature field, in which the viscous dissipation and
compression work are consistently considered.
\par

\subsubsection{Energy conservation equation}
The collision term of an LB equation conserves macroscopic quantity, and the
recovered macroscopic conservation equation for this quantity usually has a
conservative form (see Eq.\ (\ref{eq.NSE}) as an example). On the basis of
this principle, the total energy conservation equation, in which the viscous
dissipation and compression work are expressed as $\nabla \cdot (\mathbf{u}
\cdot \mathbf{\Pi}) - \nabla \cdot (p_\EOS^{} \mathbf{u})$, is a better and
more natural starting point for directly developing an LB equation at the
discrete level than the internal energy conservation equation, in which the
viscous dissipation and compression work are expressed as $\mathbf{\Pi} :
\nabla \mathbf{u} - p_\EOS^{} \nabla \cdot \mathbf{u}$. Here, $p_\EOS^{}$ is
the pressure determined by the adopted EOS. To facilitate the development of
an LB equation, the total energy conservation equation is reformulated as
\begin{equation}\label{eq.ECE}
\partial_t (\rho E) + \nabla \cdot (\rho H \mathbf{u}) =
\nabla \cdot (\lambda \nabla T + \mathbf{u} \cdot \mathbf{\Pi}) +
\mathbf{u} \cdot \mathbf{F} + q,
\end{equation}
where $E$ is the total energy, $H = E + p_\EOS^{} / \rho$ is the total
enthalpy, $T$ is the temperature that can be determined by the internal
energy $\epsilon$ ($\epsilon = E - | \mathbf{u} |^2/2$) and density $\rho$,
$\lambda$ is the heat conductivity, and $q$ is the source term. In Eq.\
(\ref{eq.ECE}), the viscous dissipation $\nabla \cdot (\mathbf{u} \cdot
\mathbf{\Pi})$ combines with the conduction term $\nabla \cdot (\lambda
\nabla T)$ to constitute the term $\nabla \cdot (\lambda \nabla T +
\mathbf{u} \cdot \mathbf{\Pi})$, and the compression work $- \nabla \cdot
(p_\EOS^{} \mathbf{u})$ combines with the convection term $\nabla \cdot
(\rho E \mathbf{u})$ to constitute the term $\nabla \cdot (\rho H
\mathbf{u})$. Meanwhile, we can also combine the work done by force
$\mathbf{u} \cdot \mathbf{F}$ and the source term $q$ to constitute an
equivalent source term $q_e = \mathbf{u} \cdot \mathbf{F} + q$. Thus, Eq.\
(\ref{eq.ECE}) can be viewed as a general convection-diffusion equation with
source term. Here, we would like to point out that the above reformulation
is consistent with the Chapman-Enskog analysis, which means that the two
terms combined together are of the same order.
\par

\subsubsection{Viscous stress tensor}
To consider the viscous dissipation in the LB equation for total energy DF,
we first recall the recovery of viscous stress tensor by the above LB
equation for density DF. On the basis of the Chapman-Enskog analysis, the
viscous stress tensor $\mathbf{\Pi}$ is of order $\varepsilon^1$ and can be
expressed as \cite{Huang2018.eos}
\begin{equation}\label{eq.Pi}
\mathbf{\Pi} = \varepsilon^1 \mathbf{\Pi}^{(1)} = - \varepsilon^1 c^2
\begin{bmatrix}
\tfrac12 \tG_7^{(1)} & \tG_8^{(1)} \\
\tG_8^{(1)}          & -\tfrac12 \tG_7^{(1)} \\
\end{bmatrix}
- \varepsilon^1 c^2
\begin{bmatrix}
\tfrac16 \tG_1^{(1)} & 0 \\
0                    & \tfrac16 \tG_1^{(1)} \\
\end{bmatrix},
\end{equation}
where $\varepsilon$ is the small expansion parameter in the Chapman-Enskog
analysis and $\tbG^{(1)}$ is
\begin{equation}\label{eq.tG}
\tbG^{(1)} = \dfrac{ \mathbf{m}^{(1)} + \bbm^{(1)} }{2} ,
\end{equation}
where $\mathbf{m} ^{(1)}$ and $\bbm^{(1)}$ are the $\varepsilon^1
\text{-order}$ terms of $\mathbf{m}$ and $\bbm$ in their Chapman-Enskog
expansions $\mathbf{m} = \sum\nolimits _{n=0}^{+\infty} \varepsilon^n
\mathbf{m} ^{(n)}$ and $\bbm = \sum\nolimits _{n=0}^{+\infty} \varepsilon^n
\bbm^{(n)}$, respectively. Here, it is worth pointing out that the
post-collision moment $\bbm^{(1)}$ is kept in Eq.\ (\ref{eq.tG}) rather than
being substituted by Eq.\ (\ref{eq.LBE.u.c}). As a consequence, the
post-collision moment $\bbm$, which is computed in the collision process of
density DF, can be directly utilized to consider the viscous dissipation in
the LB equation for total energy DF (see \ref{app.Implementation}).
Moreover, from the Chapman-Enskog analysis of the LB equation for density
DF, we can easily know that the $\varepsilon^0 \text{-order}$ terms of
$\mathbf{m}$ and $\bbm$ satisfy
\begin{equation}\label{eq.m.order1}
\mathbf{m}^{(0)} = \bar{\mathbf{m}}^{(0)} = \mathbf{m} ^\eq.
\end{equation}
\par

\subsubsection{LB equation}
For the energy conservation equation given by Eq.\ (\ref{eq.ECE}), the total
energy DF $g_i (\mathbf{x}, t)$ is introduced here, and the MRT LB equation
for $g_i (\mathbf{x}, t)$ is devised as
\begin{subequations}\label{eq.LBE.T}
\begin{equation}\label{eq.LBE.T.s}
g_i (\mathbf{x} + \mathbf{e}_i \delta_t, t + \delta_t) =
\bar{g} _i (\mathbf{x}, t),
\end{equation}
\begin{equation}\label{eq.LBE.T.c}
\bar{\mathbf{n}} (\mathbf{x}, t) = \mathbf{n} + \delta_t \mathbf{Q}_m -
\mathbf{L} \left( \mathbf{n} - \mathbf{n}^\eq +
\dfrac{\delta_t}{2} \mathbf{Q}_m \right) + c^2 \mathbf{Y} \left(
\dfrac{\mathbf{m} + \bar{\mathbf{m}}}{2} - \mathbf{m}^\eq \right),
\end{equation}
\end{subequations}
where Eqs.\ (\ref{eq.LBE.T.s}) and (\ref{eq.LBE.T.c}) represent the
streaming process in velocity space and the collision process in moment
space, respectively. The moment of total energy DF in Eq.\
(\ref{eq.LBE.T.c}) is given as $\mathbf{n} = \mathbf{M} (g_i) ^\T$, the
post-collision total energy DF in Eq.\ (\ref{eq.LBE.T.s}) is obtained via
$(\bar{g}_i) ^\T = \mathbf{M} ^{-1} \bar{\mathbf{n}}$, and the
post-collision moment $\bar{\mathbf{n}}$ is computed by Eq.\
(\ref{eq.LBE.T.c}). Here, the dimensionless transformation matrix
$\mathbf{M}$ is also given by Eq.\ (\ref{eq.M}). On the RHS of Eq.\
(\ref{eq.LBE.T.c}), the last density-DF-related term is introduced to
consider the viscous dissipation, in which $\mathbf{Y}$ is a $9 \times 9$
matrix that will be discussed and determined later. By definition, the
macroscopic total energy $\rho E$ is given as
\begin{equation}\label{eq.E}
\rho E = \sum\nolimits_i g_i + \dfrac{\delta_t}{2} q_e,
\end{equation}
where $q_e = \mathbf{u} \cdot \mathbf{F} + q$ is the equivalent source term.
Then, the total enthalpy $H$ and the temperature $T$ can be determined via
the thermodynamic relations $H = E + p_\EOS^{} / \rho$ and $T = T (\epsilon,
\rho)$ (a function of internal energy $\epsilon = E - |\mathbf{u}| ^2/2$ and
density $\rho$), respectively. In the present work, a simple relation $T =
\epsilon / C_v$, though it strictly holds only for the ideal gases, is
adopted for the sake of simplicity, and more general or empirical relations
can be adopted as required by specific applications. Here, $C_v$ is the
specific heat at constant volume.
\par

To recover the targeted energy conservation equation, as well as inspired by
the ideas of our previous works on solid-liquid phase change
\cite{Huang2015, Huang2016.amr}, the equilibrium moment function for total
energy DF $\mathbf{n} ^\eq$ is devised as
\begin{equation}
\begin{split}
\mathbf{n} ^\eq = \big[
\rho E ,\,
-4 (\rho E - \rhor C_{p,0} T) + \gamma_1^{} \rhor C_{p,0} T ,\,
4  (\rho E - \rhor C_{p,0} T) + \gamma_2^{} \rhor C_{p,0} T &,\, \\
\rho H \hu_x ,\,
-\rho H \hu_x ,\,
\rho H \hu_y ,\,
-\rho H \hu_y &,\,
0 ,\,
0
\big] ^\T,
\end{split}
\end{equation}
where $\rhor$ and $C_{p,0}$ are the reference density and the reference
specific heat at constant pressure, respectively, and $\gamma_1^{}$ and
$\gamma_2^{}$ are the coefficients related to the heat conductivity.
Similarly to $\mathbf{n} ^\eq$, the discrete source term in moment space
$\mathbf{Q}_m$ is devised as
\begin{equation}
\mathbf{Q}_m = \big(
q_e ,\,
\gamma_1^{} q_e ,\,
\gamma_2^{} q_e ,\,
q_e \hu_x ,\,
-q_e \hu_x ,\,
q_e \hu_y ,\,
-q_e \hu_y ,\,
0 ,\,
0
\big) ^\T.
\end{equation}
To avoid the deviation term caused by the convection term recovered at the
order of $\varepsilon^1$ in the diffusion term recovered at the order of
$\varepsilon^2$, the collision matrix in moment space $\mathbf{L}$ is
modified as follows \cite{Huang2014.cde}
\begin{equation}
\mathbf{L} =
\begin{bmatrix}
\sigma_0^{}& 0& 0& 0& 0& 0& 0& 0& 0\\
0& \sigma_e^{}& 0& 0& 0& 0& 0& 0& 0\\
0& 0& \sigma_\varepsilon^{}& 0& 0& 0& 0& 0& 0\\
0& 0& 0& \sigma_j^{}& \sigma_q^{} \omega_j^{}& 0& 0& 0& 0\\
0& 0& 0& 0& \sigma_q^{}& 0& 0& 0& 0\\
0& 0& 0& 0& 0& \sigma_j^{}& \sigma_q^{} \omega_j^{}& 0& 0\\
0& 0& 0& 0& 0& 0& \sigma_q^{}& 0& 0\\
0& 0& 0& 0& 0& 0& 0& \sigma_p^{}& 0\\
0& 0& 0& 0& 0& 0& 0& 0& \sigma_p^{}\\
\end{bmatrix},
\end{equation}
where $\omega_j^{} = \sigma_j^{} /2 -1$.
\par

Since the viscous stress tensor $\mathbf{\Pi}$ is only related to
$\tG_1^{(1)}$, $\tG_7^{(1)}$, and $\tG_8^{(1)}$ (see Eq.\ (\ref{eq.Pi})),
the matrix $\mathbf{Y}$ in the density-DF-related term, which is introduced
in Eq.\ (\ref{eq.LBE.T.c}) to consider the viscous dissipation, is set as
follows
\begin{equation}
\mathbf{Y} =
\begin{bmatrix}
0& 0&      0& 0& 0& 0& 0& 0&      0\\
0& 0&      0& 0& 0& 0& 0& 0&      0\\
0& 0&      0& 0& 0& 0& 0& 0&      0\\
0& Y_{31}& 0& 0& 0& 0& 0& Y_{37}& Y_{38}\\
0& Y_{41}& 0& 0& 0& 0& 0& Y_{47}& Y_{48}\\
0& Y_{51}& 0& 0& 0& 0& 0& Y_{57}& Y_{58}\\
0& Y_{61}& 0& 0& 0& 0& 0& Y_{67}& Y_{68}\\
0& 0&      0& 0& 0& 0& 0& 0&      0\\
0& 0&      0& 0& 0& 0& 0& 0&      0\\
\end{bmatrix},
\end{equation}
where $Y_{3 \alpha} + Y_{4 \alpha} = 0$ and $Y_{5 \alpha} + Y_{6 \alpha} =
0$ for $\alpha = 1$, $7$, and $8$. Through the Chapman-Enskog analysis (see
\ref{app.CE}), the nonzero elements in $\mathbf{Y}$ can be determined as
follows
\begin{equation}\label{eq.Y.nonzero}
\begin{array}{lll}
Y_{31} = \hu_x /3, & Y_{37} = \hu_x, & Y_{38} = 2 \hu_y, \\
Y_{51} = \hu_y /3, & Y_{57} =-\hu_y, & Y_{58} = 2 \hu_x. \\
\end{array}
\end{equation}
Then, the following macroscopic conservation equation can be recovered
\begin{equation}\label{eq.ECE.LBE}
\partial_t (\rho E) + \nabla \cdot (\rho H \mathbf{u}) = \nabla \cdot
\left[
\left( \dfrac{2}{3} + \dfrac{\gamma_1^{}}{2} + \dfrac{\gamma_2^{}}{3}
\right)
\rhor C_{p,0} c^2 \delta_t \left( \dfrac{1}{\sigma_j^{}} - \dfrac12
\right)
\nabla T + \mathbf{u} \cdot \mathbf{\Pi} \right] + q_e.
\end{equation}
Compared with Eq.\ (\ref{eq.ECE}), the heat conductivity is given as
$\lambda = (2/3 + \gamma_1^{}/2 + \gamma_2^{}/3) \rhor C_{p,0} c^2 \delta_t
\big(\sigma_j^{-1} - 0.5\big)$. It can be seen from Eq.\ (\ref{eq.ECE.LBE})
that the viscous dissipation and compression work are correctly considered.
\par

Before proceeding further, some discussion on the present LB model for
coupled thermo-hydrodynamic flows is in order. First, the MRT collision
scheme is employed in both the LB equations for density and total energy
DFs, and the collision matrix in moment space is modified to be a
nondiagonal matrix rather than being set as the conventional diagonal
matrix. Second, the Prandtl number $\Prt = C_p \mu / \lambda$ can be
arbitrarily adjusted. Here, $C_p$ is the specific heat at constant pressure,
and $\mu = \rho \nu$ is the dynamic viscosity. Third, the specific heat
ratio $\gamma = C_p / C_v$ can also be arbitrarily adjusted. Note that $C_p
- C_v$ depends on the adopted EOS, and $C_p - C_v = R_g$ holds only for the
ideal-gas EOS. Lastly, and most importantly, an arbitrary EOS (including the
nonideal-gas EOS) $p_\EOS^{}$ can be prescribed, and the built-in variable
$\eta$ is inversely calculated via $\eta = [ c_s^{-2} p_\LBE^{} - (2 +
\alpha_1^{}) \rho ] \big/ \beta_1^{}$ with $p_\LBE^{} = p_\EOS^{}$.
\par

\section{Boundary condition treatment} \label{sec.bc}
In real applications, the boundary conditions are usually given in terms of
the macroscopic variables, and thus additional treatment is required to
obtain the mesoscopic DFs at the boundary node. In this section, we propose
the boundary condition treatment for simulating coupled thermo-hydrodynamic
flows.
\par

\subsection{Macroscopic variables}
\begin{figure}[htbp]
  \centering
  \includegraphics[scale=1,draft=\figdraft]{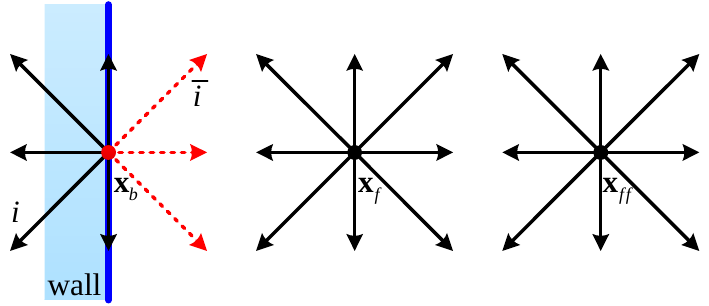}
  \caption{Schematic of boundary condition treatment with $\mathbf{x}_b$
  denoting the boundary node placed on the wall, $\mathbf{x}_f$ and
  $\mathbf{x}_\ff$ denoting the nearest and next-nearest fluid nodes in the
  normal direction, and the solid and dashed arrow lines denoting the known
  and unknown distribution functions after the streaming process.}
  \label{fig.BC}
\end{figure}

For the velocity field, the nonslip velocity boundary condition is
considered and the velocity on the boundary is directly specified. Due to
the full coupling of thermo-hydrodynamic effects, the density may
significantly vary near the boundary and also has a direct effect on the
heat transfer process. Thus, it is important to ensure the mass conservation
at the boundary node for simulating coupled thermo-hydrodynamic flows. In
the present boundary condition treatment, the boundary node $\mathbf{x}_b$
is exactly placed on the wall boundary, as shown in Fig.\ \ref{fig.BC}. The
post-collision density DF $\bar{f}_i (\mathbf{x}_b, t)$ hitting the wall
(i.e., streaming out of the computational domain) reverses its direction as
follows
\begin{equation}\label{eq.BB.f}
f_{\bar{i}, \text{temp}} (\mathbf{x}_b, t + \delta_t) = \bar{f}_i
(\mathbf{x}_b, t),
\end{equation}
where $\bar{i}$ means $\mathbf{e}_{\bar{i}} = - \mathbf{e}_i$, and the
subscript ``temp'' implies that the density DF $f_{\bar{i}, \text{temp}}
(\mathbf{x}_b, t + \delta_t)$ is temporary. After this ``bounce-back''
process, all the unknown density DFs at $\mathbf{x}_b$ and $t + \delta_t$
due to the absence of adjacent nodes are now obtained. Then, the density
$\rho (\mathbf{x}_b, t+\delta_t)$ can be computed via definition as usual
(i.e., $\rho = \sum\nolimits_i f_i$). Note that the velocity $\mathbf{u}
(\mathbf{x}_b, t + \delta_t)$ is directly specified. Obviously, the local
conservation of mass can be strictly satisfied at the boundary node.
\par

As for the temperature field, the Dirichlet boundary condition with
specified temperature and the Neumann boundary condition with zero heat flux
(i.e., the adiabatic boundary condition) are considered. For the Dirichlet
boundary condition, since the temperature $T (\mathbf{x}_b, t + \delta_t)$
is directly specified, all the involved macroscopic variables, such as the
total energy $E (\mathbf{x}_b, t + \delta_t)$, the pressure $p_\EOS^{}
(\mathbf{x}_b, t + \delta_t)$, and the total enthalpy $H (\mathbf{x}_b, t +
\delta_t)$, can be determined via the corresponding thermodynamic relations.
For the Neumann boundary condition with zero heat flux, the post-collision
total energy DF $\bar{g}_i (\mathbf{x}_b, t)$ hitting the wall reverses its
direction as follows
\begin{equation}\label{eq.BB.g}
g_{\bar{i}, \text{temp}} (\mathbf{x}_b, t + \delta_t) = \bar{g}_i
(\mathbf{x}_b, t),
\end{equation}
and thus all the unknown total enthalpy DFs at $\mathbf{x}_b$ and $t +
\delta_t$ due to the absence of adjacent nodes are temporarily obtained.
Then, the total energy $E (\mathbf{x}_b, t + \delta_t)$ can be computed via
definition as usual (i.e., $\rho E = \sum\nolimits_i g_i + \delta_t q_e
/2$), and all the involved macroscopic variables, such as the temperature $T
(\mathbf{x}_b, t + \delta_t)$, the pressure $p_\EOS^{} (\mathbf{x}_b, t +
\delta_t)$, and the total enthalpy $H (\mathbf{x}_b, t + \delta_t)$, can be
determined via the corresponding thermodynamic relations.
\par

\subsection{Density and total energy DFs}
At the boundary node, the unknown density and total energy DFs obtained via
Eqs.\ (\ref{eq.BB.f}) and (\ref{eq.BB.g}) are only used to compute the
macroscopic density and total energy. In the present boundary condition
treatment, all the known and unknown DFs at the boundary node will be
updated to make sure that the defining equations of density, velocity, and
total energy (i.e., Eqs.\ (\ref{eq.rho.u}) and (\ref{eq.E})) exactly hold at
the boundary node. For this purpose, we decompose the moment of DF (the same
as the DF) into its equilibrium, force (source), and nonequilibrium parts,
i.e.,
\begin{subequations}\label{eq.decompose}
\begin{equation}
\mathbf{m} = \mathbf{m}^\eq - \dfrac{\delta_t}{2} \mathbf{F}_m +
\mathbf{m} ^\text{neq},
\end{equation}
\begin{equation}
\mathbf{n} = \mathbf{n}^\eq - \dfrac{\delta_t}{2} \mathbf{Q}_m +
\mathbf{n} ^\text{neq}.
\end{equation}
\end{subequations}
Note that the present nonequilibrium part $\mathbf{m}^\text{neq}$
($\mathbf{n}^\text{neq}$) in Eq.\ (\ref{eq.decompose}) is different from the
previous nonequilibrium part defined as $\mathbf{m} - \mathbf{m}^\eq$
($\mathbf{n} - \mathbf{n}^\eq$) \cite{Kruger2017} when the force (source)
term exists. Since the equilibrium parts ($\mathbf{m}^\eq$ and
$\mathbf{n}^\eq$) and the force (source) parts ($- \delta_t \mathbf{F}_m /2$
and $-\delta_t \mathbf{Q}_m /2$) are determined by the macroscopic
variables, $\mathbf{m}^\eq (\mathbf{x}_b, t+\delta_t) - \delta_t
\mathbf{F}_m (\mathbf{x}_b, t+\delta_t) /2$ and $\mathbf{n}^\eq
(\mathbf{x}_b, t+\delta_t) - \delta_t \mathbf{Q}_m (\mathbf{x}_b,
t+\delta_t) /2$ can be directly computed. As to the nonequilibrium parts
($\mathbf{m} ^\text{neq}$ and $\mathbf{n} ^\text{neq}$) at $\mathbf{x}_b$
and $t + \delta_t$, extrapolations are employed following the idea of the
nonequilibrium-extrapolation approach \cite{Guo2002a, Guo2002b}. However,
instead of simply extrapolating $\mathbf{m} ^\text{neq}$ and $\mathbf{n}
^\text{neq}$, we introduce the following terms
\begin{subequations}\label{eq.neq.redefinition}
\begin{equation}
\tilde{\mathbf{m}} ^\text{neq} = \left( \mathbf{I} -
\dfrac{\mathbf{S}}{2}
\right) \mathbf{m} ^\text{neq},
\end{equation}
\begin{equation}
\tilde{\mathbf{n}} ^\text{neq} = \left( \mathbf{I} -
\dfrac{\mathbf{L}}{2}
\right) \mathbf{n} ^\text{neq} + \dfrac{c^2 \mathbf{Y}}{2} \left(
\mathbf{I}
- \dfrac{\mathbf{S}}{2} \right) \mathbf{m} ^\text{neq},
\end{equation}
\end{subequations}
where $\mathbf{I}$ is the $9 \times 9$ identity matrix; and then the first-
and second-order nonequilibrium extrapolations are given as
\begin{subequations}\label{eq.neq.extrapolation}
\begin{equation}
\begin{array}{ll}
\text{first-order:} & \tilde{\mathbf{m}} ^\text{neq} (\mathbf{x}_b, t +
\delta_t) = \tilde{\mathbf{m}} ^\text{neq} (\mathbf{x}_f, t +
\delta_t), \\
\text{second-order:} & \tilde{\mathbf{m}} ^\text{neq} (\mathbf{x}_b, t +
\delta_t) = 2 \tilde{\mathbf{m}} ^\text{neq} (\mathbf{x}_f, t +
\delta_t) - \tilde{\mathbf{m}} ^\text{neq} (\mathbf{x}_\ff, t +
\delta_t);
\end{array}
\end{equation}
\begin{equation}
\begin{array}{ll}
\text{first-order:} & \tilde{\mathbf{n}} ^\text{neq} (\mathbf{x}_b, t +
\delta_t) = \tilde{\mathbf{n}} ^\text{neq} (\mathbf{x}_f, t +
\delta_t), \\
\text{second-order:} & \tilde{\mathbf{n}} ^\text{neq} (\mathbf{x}_b, t +
\delta_t) = 2 \tilde{\mathbf{n}} ^\text{neq} (\mathbf{x}_f, t +
\delta_t) - \tilde{\mathbf{n}} ^\text{neq} (\mathbf{x}_\ff, t +
\delta_t);
\end{array}
\end{equation}
\end{subequations}
where $\mathbf{x}_f$ and $\mathbf{x}_\ff$ denote the nearest and
next-nearest fluid nodes in the normal direction, as shown in Fig.\
\ref{fig.BC}. Based on our numerical tests, the first-order extrapolation
has better stability but lower accuracy than the second-order extrapolation.
Note that although the present collision matrices $\mathbf{S}$ and
$\mathbf{L}$ are nondiagonal, $\mathbf{I} - \mathbf{S}/2$ and $\mathbf{I} -
\mathbf{L}/2$ in Eq.\ (\ref{eq.neq.redefinition}) are still invertible, and
their inverse matrices are given in \ref{app.Inverse}. Therefore, Eq.\
(\ref{eq.neq.extrapolation}) is compatible and can be easily implemented due
to the special forms of $( \mathbf{I} - \mathbf{S}/2 )^{-1}$ and $(
\mathbf{I} - \mathbf{L}/2 )^{-1}$. Moreover, different from the previous
nonequilibrium-extrapolation approach \cite{Guo2002a, Guo2002b, Guo2007},
the present boundary condition treatment is applicable to the situation when
the dynamic viscosity and thermal conductivity significantly vary with
temperature and hence with space because the collision matrices are
considered in the present extrapolations of nonequilibrium parts.
\par

\section{Validations and discussions} \label{sec.validation}
In this section, simulations of thermal Poiseuille and Couette flows are
first carried out to validate the present LB model with self-tuning EOS for
coupled thermo-hydrodynamic flows. Three different EOSs, including the
decoupling EOS, the ideal-gas EOS, and the Carnahan-Starling EOS for
rigid-sphere fluids \cite{Carnahan1969}, are adopted, which are given in
order as follows
\begin{subequations}\label{eq.EOSs}
\begin{equation}\label{eq.EOSs.1}
p_\EOS^{} = \rho R_g T_0,
\end{equation}
\begin{equation}\label{eq.EOSs.2}
p_\EOS^{} = \rho R_g T,
\end{equation}
\begin{equation}\label{eq.EOSs.3}
p_\EOS^{} = Z \rho R_g T \quad \text{with} \quad Z= \dfrac{1 + b\rho/4 +
(b\rho/4)^2 - (b\rho/4)^3}{ (1-b\rho/4)^3 },
\end{equation}
\end{subequations}
where $Z$ is the compressibility factor with the coefficient $b$ set to
$\sqrt{2} \pi / (3 \rhor)$ here. Then, the present LB model is applied to
the simulation of natural convection in a square cavity with a large
temperature difference. The ideal-gas EOS is adopted and the Rayleigh number
varies from $10^3$ up to $10^8$. In the following simulations, $\varpi =
1/6$, $\gamma_1^{} = -2$, and $\gamma_2^{} = 2$ are chosen. The relaxation
parameters in $\mathbf{S}$ satisfy $s_0^{} = s_j^{} =1$, $\big( s_p^{-1} -
0.5 \big) \big( s_q^{-1} - 0.5 \big) = 1/12$, and $s_\varepsilon^{} =
s_e^{}$ \cite{Huang2016.3rd}, and the relaxation parameters in $\mathbf{L}$
satisfy $\sigma_0^{} =1$, $\big( \sigma_j^{-1} - 0.5 \big) \big(
\sigma_e^{-1} - 0.5 \big) = 1/4$, $\sigma_\varepsilon^{} = \sigma_p^{} =
\sigma_e^{}$, and $\sigma_q^{} = \sigma_j^{}$ \cite{Huang2016.amr}.
Meanwhile, the ratio of bulk to kinematic viscosity $\varsigma / \nu$ is
fixed at $1$ unless otherwise stated.
\par

\subsection{Thermal Poiseuille flow}
The thermal Poiseuille flow, driven by a constant force $\mathbf{F} = (F_x,
\, 0) ^\T$ between two parallel walls, is first simulated. Both the lower
and upper walls are at rest, and the temperature of the lower and upper
walls are kept at $T_c$ and $T_h$ ($T_c < T_h$), respectively. The Prandtl
number $\Prt = C_p \mu / \lambda$, the specific heat at constant pressure
$C_p$, and the dynamic viscosity $\mu$ are assumed to be constant. Thus, the
analytical solutions for velocity and temperature are given as
\cite{Guo2007}
\begin{subequations}\label{eq.Poiseuille.u.T}
\begin{equation}
\dfrac{u_x}{U_0} = 4 \dfrac{y}{W} \left( 1- \dfrac{y}{W} \right), \quad
\dfrac{u_y}{U_0} =0,
\end{equation}
\begin{equation}
\dfrac{T - T_c}{T_h - T_c} = \dfrac{\Prt \Ec}{3} \left[ 1 - \left( 1 -
2\dfrac{y}{W} \right)^4 \right] + \dfrac{y}{W},
\end{equation}
\end{subequations}
where $W$ is the channel width, $U_0 = F_x W^2 / (8 \mu)$ is the maximum
velocity, and $\Ec = U_0^2 \big/ [C_p (T_h - T_c)]$ is the Eckert number. As
seen in Eq.\ (\ref{eq.Poiseuille.u.T}), the analytical solutions for
velocity and temperature are fully determined by $\Prt$ and $\Ec$. However,
the analytical solution for density further depends on both the initial
state and the adopted EOS. In the simulations, the density, velocity, and
temperature are initialized as $\rhor$, $\mathbf{0}$, and $T_0$ ($T_0 =
T_c$), respectively, and the initial pressure is determined by the adopted
EOS. Thus, for the decoupling EOS (i.e., Eq.\ (\ref{eq.EOSs.1})), the
analytical solution for density can be easily obtained as
\begin{subequations}\label{eq.rho.analytical}
\begin{equation}\label{eq.rho.analytical.1}
\dfrac{\rho}{\rhor} = 1;
\end{equation}
for the ideal-gas EOS (i.e., Eq.\ (\ref{eq.EOSs.2})), the analytical
solution for density is given as
\begin{equation}\label{eq.rho.analytical.2}
\dfrac{\rho}{\rhor} = A \dfrac{T_0}{T},
\end{equation}
where the coefficient $A^{-1} = \int\nolimits_0^W T_0 / T \text{d}y$; as for
the Carnahan-Starling EOS (i.e., Eq.\ (\ref{eq.EOSs.3})), the analytical
solution for density satisfies
\begin{equation}\label{eq.rho.analytical.3}
\int\limits_0^W \rho \text{d}y = \rhor W \quad \text{and} \quad
p_\infty^{} = Z \rho R_g T,
\end{equation}
\end{subequations}
where $p_\infty^{}$ is the final pressure in the channel. Although an
explicit expression for $\rho / \rhor$ cannot be derived from Eq.\
(\ref{eq.rho.analytical.3}), $\rho / \rhor$ can be easily obtained with high
precision using numerical integration.
\par

In the simulations, the lattice sound speed is set as
\begin{equation}\label{eq.cs}
c_s = \left. \sqrt{\partial_\rho p_\EOS^{}} \right|
\lower 1.23ex \hbox{$\scriptstyle \rho = \rhor, T = T_0$},
\end{equation}
and the specific heats at constant pressure and volume are fixed at
\begin{equation}\label{eq.Cp}
C_p = 3.5 \dfrac{c_s^2}{T_0}, \quad C_v = C_p - \dfrac{p_\EOS^{}
(\rhor, T_0)}{\rhor T_0}.
\end{equation}
With this configuration, the specific heat ratio $\gamma$ is $1.4$ for the
decoupling and ideal-gas EOSs and $1.101465$ for the Carnahan-Starling EOS.
The simulations are carried out on a $64 \times 64$ grid with lattice
spacing $\delta_x = 1/64$ and periodic boundary in $x \text{-direction}$.
The lower and upper walls are treated by the present boundary condition
treatment with second-order extrapolation. The basic parameters are set as
$R_g = 1$, $T_0 = 1$, and $\rhor = 1$, and the dimensionless relaxation time
for density DF, defined as $\tau = s_p^{-1}$, is fixed at $0.8$ for the
kinematic viscosity $\mu / \rhor$.
\par

\begin{figure}[htbp]
  \centering
  \includegraphics[scale=1,draft=\figdraft]{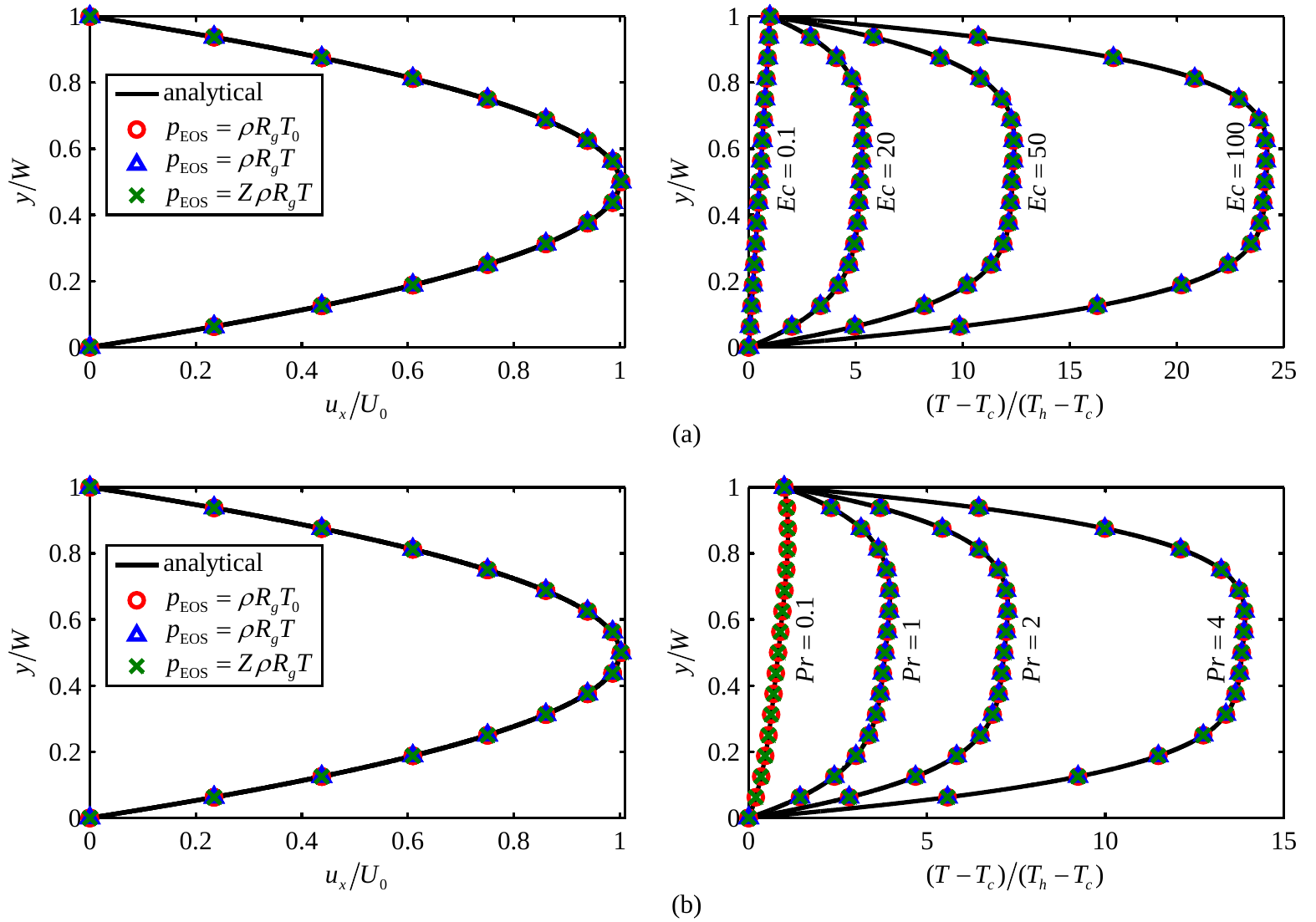}
  \caption{Comparisons of velocity $u_x / U_0$ (left) and temperature
  $(T - T_c) / (T_h - T_c)$ (right) distributions between the numerical
  results and the analytical solutions for thermal Poiseuille flow with
  (a) $\Prt = 0.71$ and $\Ec$ varying from $0.1$ to $100$, and
  (b) $\Ec = 10$ and $\Prt$ varying from $0.1$ to $4$.}
  \label{fig.Poiseuille.u.T}
\end{figure}

\begin{figure}[htbp]
  \centering
  \includegraphics[scale=1,draft=\figdraft]{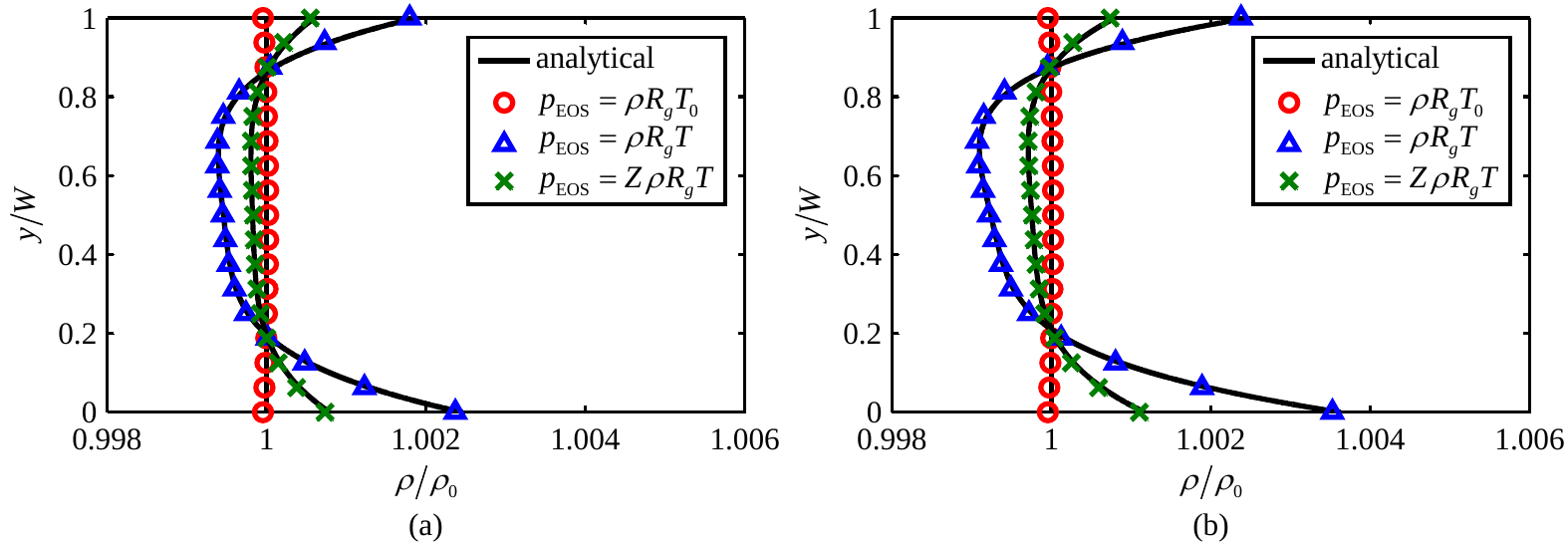}
  \caption{Comparisons of density $\rho / \rhor$ distributions between
  the numerical results obtained with different EOSs and the corresponding
  analytical solutions for thermal Poiseuille flow with
  (a) $\Prt = 0.71$ and $\Ec = 20$, and (b) $\Ec = 10$ and $\Prt=1$.}
  \label{fig.Poiseuille.rho}
\end{figure}

Fig.\ \ref{fig.Poiseuille.u.T} shows the velocity $u_x / U_0$ and
temperature $(T - T_c) / (T_h - T_c)$ distributions across the channel and
compares the numerical results with the analytical solutions given by Eq.\
(\ref{eq.Poiseuille.u.T}). Two sets of $\Prt$ and $\Ec$ are considered here:
for the first set, $\Prt$ is fixed at $0.71$ and $\Ec$ varies from $0.1$ to
$100$; while for the second set, $\Ec$ is fixed at $10$ and $\Prt$ varies
from $0.1$ to $4$. As an important computational parameter, the lattice Mach
number, defined as $\Ma_\text{lattice} = U_0 / c_s$, is fixed at $0.2$ in
the simulations. Good agreement between the numerical results and the
analytical solutions can be observed in Fig.\ \ref{fig.Poiseuille.u.T},
which demonstrates that the effects of the viscous dissipation and
compression work are successfully captured by the present LB model. From
Fig.\ \ref{fig.Poiseuille.u.T}, we can also see that the distributions of
$u_x / U_0$ and $(T - T_c) / (T_h - T_c)$ obtained with different EOSs are
almost identical, which agrees with the aforementioned discussion. Note that
the simulation with $\Ec = 10$ and $\Prt = 0.1$ loses numerical stability
for the ideal-gas EOS. To further validate the present LB model with
self-tuning EOS, comparisons of the density $\rho / \rhor$ distributions are
carried out in Fig.\ \ref{fig.Poiseuille.rho}. Good agreement is observed
between the numerical results obtained with different EOSs and the
corresponding analytical solutions given by Eq.\ (\ref{eq.rho.analytical}),
which demonstrates that various EOSs (including the nonideal-gas EOS) can be
handled by the present LB model. For the decoupling EOS, $\rho / \rhor$
keeps constant across the channel; as for the ideal-gas and
Carnahan-Starling EOSs, $\rho / \rhor$ varies across the channel due to the
full coupling of thermo-hydrodynamic effects. In the Carnahan-Starling EOS,
the molecular volume is considered, which implies that the rigid-sphere
fluid is less compressible than the corresponding ideal gas. Therefore, the
variation in density across the channel obtained with the Carnahan-Starling
EOS is smaller than that obtained with the ideal-gas EOS, as clearly shown
in Fig.\ \ref{fig.Poiseuille.rho}.
\par

\begin{figure}[htbp]
  \centering
  \includegraphics[scale=1,draft=\figdraft]{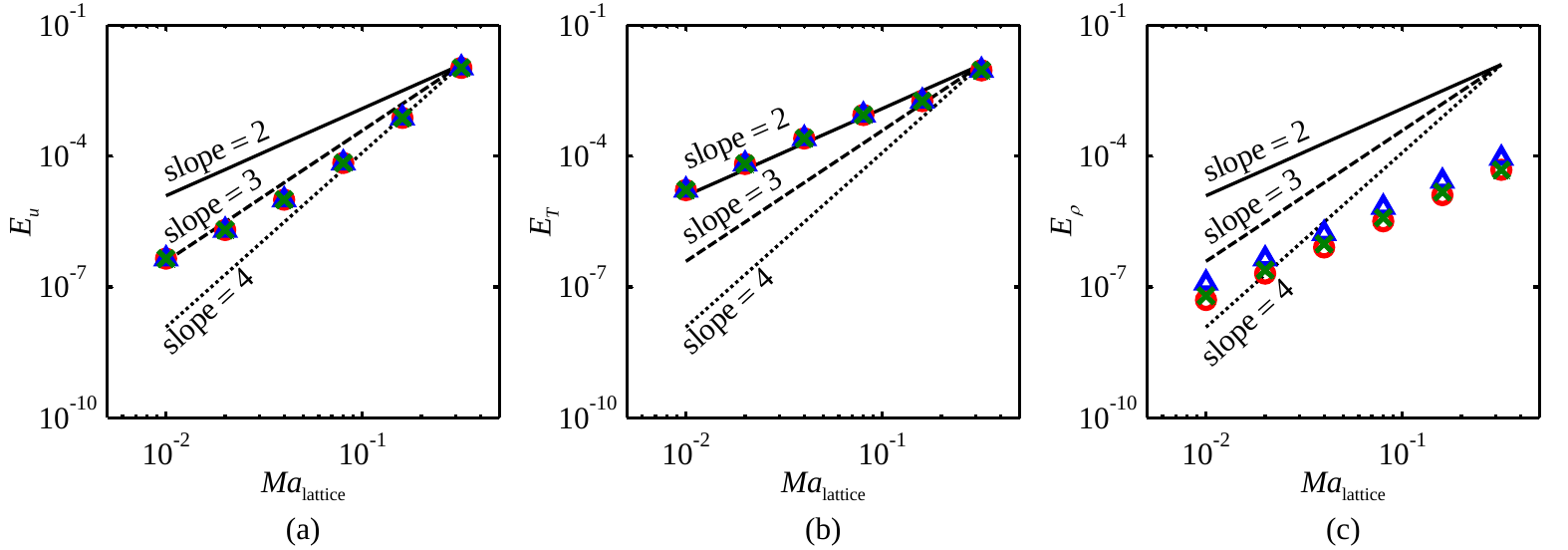}
  \caption{Relative errors of (a) velocity $u_x / U_0$, (b) temperature
  $(T - T_c) / (T_h - T_c)$, and (c) density $\rho / \rhor$ versus lattice
  Mach number $\Ma_\text{lattice}$ when $\Prt = 0.71$ and
  $\Ec = 100 \Ma _\text{lattice} ^2$. The symbols ``$\circ$'',
  ``$\vartriangle$'', and ``$\times$'' denote the results obtained with
  decoupling, ideal-gas, and Carnahan-Starling EOSs, respectively,
  and the solid, dashed, and dotted lines denote second-, third-, and
  fourth-order accuracy with respect to $\Ma _\text{lattice}$, respectively.}
  \label{fig.Poiseuille.Ma}
\end{figure}

Considering that the lattice Mach number plays an important role in the LB
method, we further investigate the accuracy of the present simulation with
respect to $\Ma_\text{lattice}$. It is worth pointing out that the lattice
Mach number is not only a computational parameter but also closely related
to the Eckert number (or the real Mach number) due to the lattice sound
speed given by Eq.\ (\ref{eq.cs}). In the following simulations, the Prandtl
number is fixed at $0.71$, and the Eckert number is set as $100 \Ma
_\text{lattice} ^2$ with $\Ma _\text{lattice}$ varying from $0.01$ to
$0.32$. Thus, the temperature difference $T_h - T_c$ remains unchanged for
different $\Ma _\text{lattice}$. The relative errors of velocity,
temperature, and density are calculated here, which are defined as
\begin{equation}\label{eq.Err}
E_\phi = \sqrt{ \dfrac{ \sum [f(\phi)_\text{numerical} -
f(\phi)_\text{analytical} ]^2}{ \sum f(\phi)_\text{analytical} ^2} },
\end{equation}
where $f(\phi)$ denotes the velocity $u_x / U_0$, temperature $(T - T_c) /
(T_h - T_c)$, and density $\rho / \rhor$ when $\phi = u$, $T$, and $\rho$,
respectively, the subscripts ``numerical'' and ``analytical'' denote the
numerical result and analytical solution of $f(\phi)$, respectively, and the
summation is over the computational domain. The relative errors $E_u$,
$E_T$, and $E_\rho$ versus $\Ma _\text{lattice}$ are shown in Fig.\
\ref{fig.Poiseuille.Ma}. As seen, the accuracy with respect to $\Ma
_\text{lattice}$ for velocity $u_x / U_0$ is fourth order when $\Ma
_\text{lattice}$ is relatively large and gradually decreases to second order
as $\Ma _\text{lattice}$ decreases, while the accuracy for temperature $(T -
T_c) / (T_h - T_c)$ and density $\rho / \rhor$ keep second order. Here, the
fourth-order accuracy for $u_x / U_0$ when $\Ma _\text{lattice}$ is
relatively large is due to the elimination of the additional cubic terms of
velocity, and the second-order accuracy may be caused by the boundary
condition treatment. Nevertheless, from Fig.\ \ref{fig.Poiseuille.Ma} we can
clearly see that satisfying results with different EOSs can be obtained by
the present LB model and boundary condition treatment under the low Mach
number condition.
\par

\subsection{Thermal Couette flow}
The thermophysical properties (dynamic viscosity and thermal conductivity)
are assumed to be constant for the above thermal Poiseuille flow. To
validate that the present LB model is capable of handling the coupled
thermo-hydrodynamic flows with variable thermophysical properties, the
thermal Couette flow between two parallel walls is simulated in this
section. The lower wall is at rest and keeps adiabatic, and the upper wall
moves along $x$-direction with a constant velocity $U_0$ and keeps at a
constant temperature $T_0$. The Prandtl number $\Prt = C_p \mu / \lambda$
and the specific heat at constant pressure $C_p$ are assumed to be constant,
and thus $\lambda \propto \mu$. Considering a linear dependence of $\mu$ on
$T$ that is $\mu / \mu_0 = T / T_0$, the analytical solutions for velocity
and temperature are given as \cite{Liepmann1957}
\begin{subequations}\label{eq.Couette.u.T}
\begin{equation}
\dfrac{u_x}{U_0} + \dfrac{\Prt \Ma_e^2}{2} \left( \dfrac{u_x}{U_0} -
\dfrac13 \dfrac{u_x^3}{U_0^3} \right) = \left( 1 +
\dfrac{\Prt \Ma_e^2}{3} \right) \dfrac{y}{W}, \quad
\dfrac{u_y}{U_0} = 0,
\end{equation}
\begin{equation}
\dfrac{T}{T_0} = 1 + \dfrac{\Prt \Ma_e^2}{2} \left( 1 -
\dfrac{u_x^2}{U_0^2} \right),
\end{equation}
\end{subequations}
where $W$ is the channel width, and $\Ma_e = U_0 / \sqrt{C_p T_0}$ is an
equivalent Mach number different from but closely related to the lattice and
real Mach numbers. As it can be seen from Eq.\ (\ref{eq.Couette.u.T}), the
analytical solutions for $u_x / U_0$ and $T / T_0$ are fully determined by
$\Prt$ and $\Ma_e$. Similarly to the thermal Poiseuille flow, the analytical
solution for density $\rho / \rhor$ here is not only related to the initial
state, but it also depends on the adopted EOS. In the simulations, the
density, velocity, temperature, and pressure are initialized as $\rhor$,
$\mathbf{0}$, $T_0$, and $p_\EOS^{} (\rhor, T_0)$, respectively. Thus, the
analytical solution for density $\rho / \rhor$ is also given by Eq.\
(\ref{eq.rho.analytical}), where the coefficient $A$ can be explicitly
written as $A = 1 + \Prt \Ma_e^2 \big/3$.
\par

\begin{figure}[htbp]
  \centering
  \includegraphics[scale=1,draft=\figdraft]{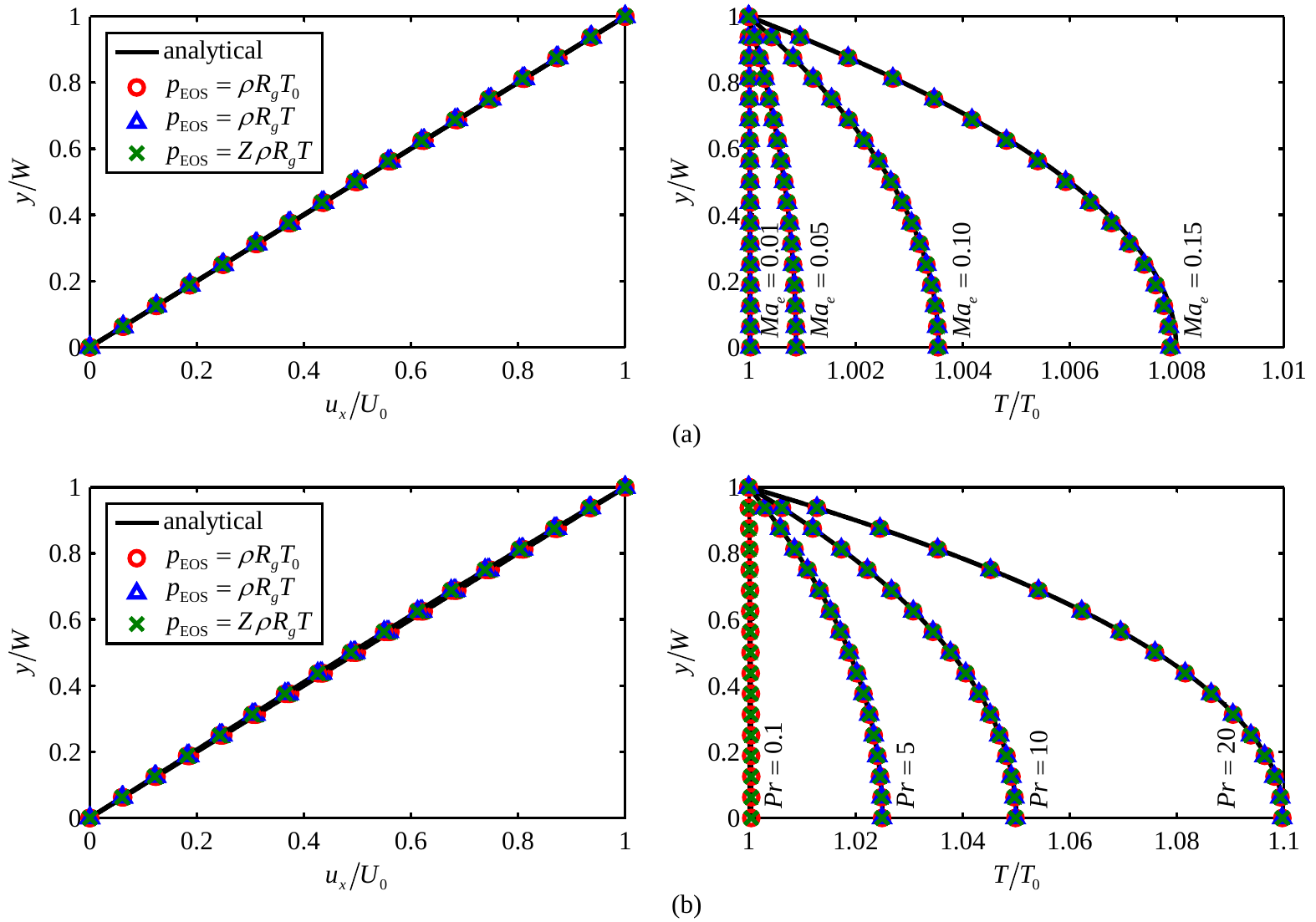}
  \caption{Comparisons of velocity $u_x / U_0$ (left) and temperature
  $T / T_0$ (right) distributions between the numerical results and the
  analytical solutions for thermal Couette flow with (a) $\Prt = 0.71$ and
  $\Ma_e$ varying from $0.01$ to $0.15$, and (b) $\Ma_e = 0.10$ and $\Prt$
  varying from $0.1$ to $20$.}
  \label{fig.Couette.u.T}
\end{figure}

\begin{figure}[htbp]
  \centering
  \includegraphics[scale=1,draft=\figdraft]{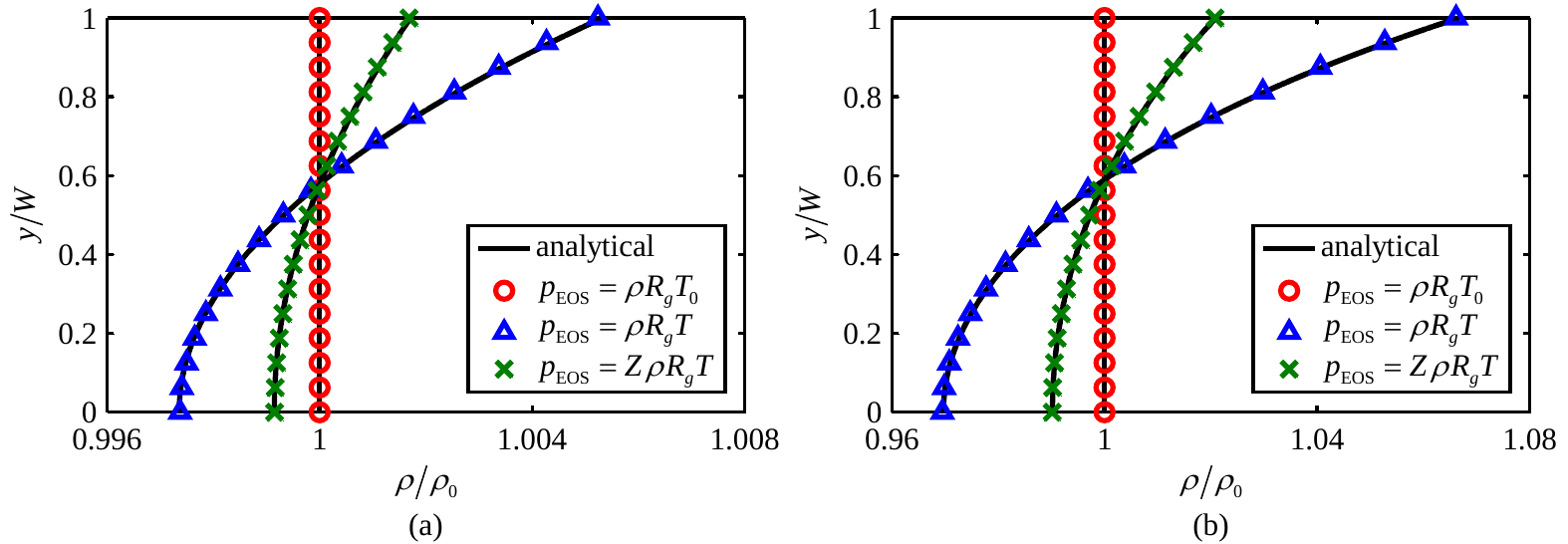}
  \caption{Comparisons of density $\rho / \rhor$ distributions between the
  numerical results obtained with different EOSs and the corresponding
  analytical solutions for thermal Couette flow with (a) $\Prt = 0.71$ and
  $\Ma_e = 0.15$, and (b) $\Ma_e = 0.10$ and $\Prt = 20$.}
  \label{fig.Couette.rho}
\end{figure}

In the simulations, all the simulation parameters are chosen the same as
those for the thermal Poiseuille flow, except that $\tau$ is fixed at $0.8$
for $\mu_0 / \rhor$. Since $\mu$ varies with $T$, $\tau$ also varies with
$T$ even for the decoupling EOS. Fig.\ \ref{fig.Couette.u.T} gives the
velocity $u_x / U_0$ and temperature $T / T_0$ distributions across the
channel for $\Prt = 0.71$ and $\Ma_e$ varying from $0.01$ to $0.15$, and for
$\Ma_e = 0.10$ and $\Prt$ varying from $0.1$ to $20$. Here, the simulation
with $\Ma_e = 0.10$ and $\Prt = 0.1$ also loses numerical stability for the
ideal-gas EOS. As seen in Fig.\ \ref{fig.Couette.u.T}, the numerical results
are in good agreement with the analytical solutions. Thus, the coupled
thermo-hydrodynamic flows with variable thermophysical properties can be
successfully handled by the present LB model. Fig.\ \ref{fig.Couette.u.T}
also verifies that the results ($u_x /U_0$ and $T / T_0$) obtained with
different EOSs are indistinguishable as long as the simulations are
numerically stable. Fig.\ \ref{fig.Couette.rho} compares the distributions
of density $\rho / \rhor$ obtained with different EOSs for $\Prt = 0.71$ and
$\Ma_e = 0.15$, and for $\Ma_e = 0.10$ and $\Prt = 20$. Good agreement
between the numerical results and the corresponding analytical solutions can
be observed, which reaffirms the applicability and accuracy of the present
LB model with self-tuning EOS.
\par

\begin{figure}[htbp]
  \centering
  \includegraphics[scale=1,draft=\figdraft]{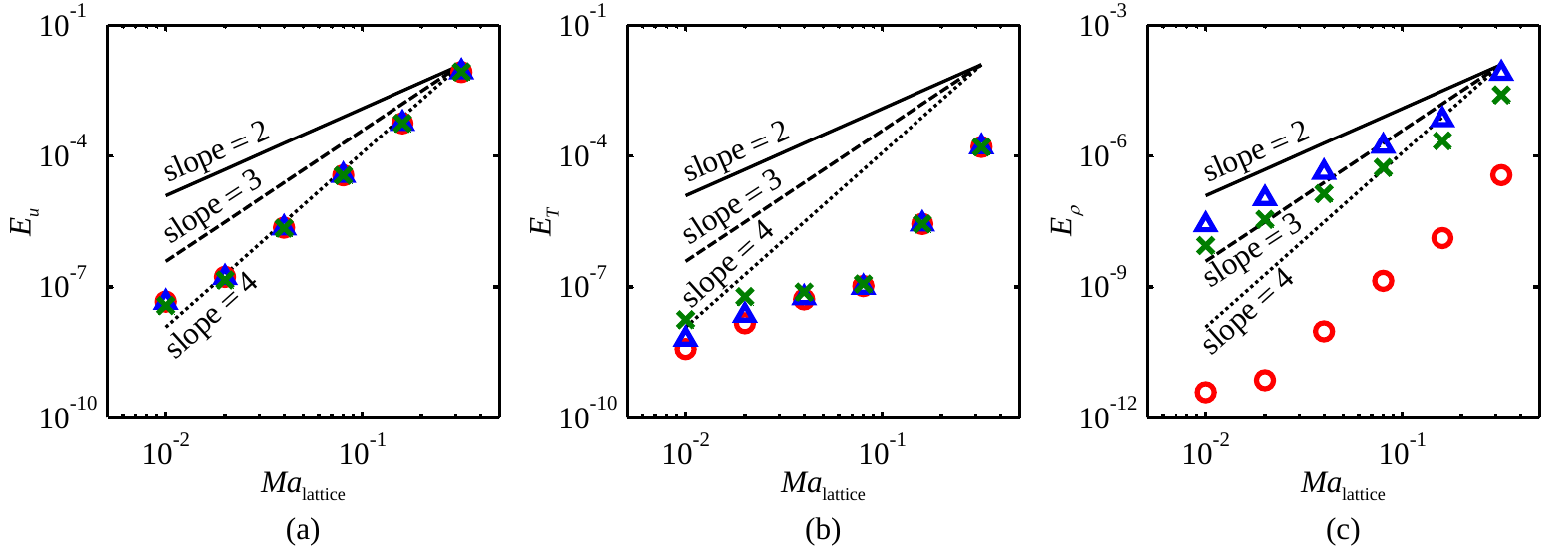}
  \caption{Relative errors of (a) velocity $u_x / U_0$, (b) temperature
  $T / T_0$, and (c) density $\rho / \rhor$ versus lattice Mach number
  $\Ma _\text{lattice}$ when $\Prt = 0.71$ and $\Ma_e = \Ma _\text{lattice}
  / \sqrt{3.5}$. The symbols ``$\circ$'', ``$\vartriangle$'', and
  ``$\times$'' denote the results obtained with decoupling, ideal-gas, and
  Carnahan-Starling EOSs, respectively, and the solid, dashed, and dotted
  lines denote second-, third-, and fourth-order accuracy with respect
  to $\Ma _\text{lattice}$, respectively.}
  \label{fig.Couette.Ma}
\end{figure}

The accuracy of the present simulation with respect to the lattice Mach
number $\Ma _\text{lattice} = U_0 / c_s$ is also investigated here.
Considering the specific heat at constant pressure given by Eq.\
(\ref{eq.Cp}), we have $\Ma_e = \Ma _\text{lattice} / \sqrt{3.5}$. In the
following simulations, the Prandtl number $\Prt$ is fixed at $0.71$, and the
lattice Mach number $\Ma _\text{lattice}$ varies from $0.01$ to $0.32$.
Fig.\ \ref{fig.Couette.Ma} shows the variations of the relative errors
$E_u$, $E_T$, and $E_\rho$ with $\Ma _\text{lattice}$. Here, the relative
error $E_\phi$ ($\phi = u$, $T$, and $\rho$) is also computed via Eq.\
(\ref{eq.Err}), in which $f(\phi)$ denotes $u_x / U_0$, $T / T_0$, and $\rho
/ \rhor$ when $\phi = u$, $T$, and $\rho$, respectively. It can be seen from
Fig.\ \ref{fig.Couette.Ma}(a) that the accuracy with respect to $\Ma
_\text{lattice}$ for velocity $u_x / U_0$ is fourth order and decreases to
second order when $\Ma _\text{lattice}$ and also $E_u$ are very small. A
similar trend can also be observed in Fig.\ \ref{fig.Couette.Ma}(b) for the
accuracy for temperature $T / T_0$. As to the accuracy for density $\rho /
\rhor$, it is fourth order and decreases rapidly when $E_\rho$ is rather
small for the decoupling EOS, while it is second order for the ideal-gas and
Carnahan-Starling EOSs. Here, the observed high-order accuracy with respect
to $\Ma _\text{lattice}$ can be explained by the elimination of the
additional cubic terms of velocity in the recovered momentum conservation
equation, and the deterioration of accuracy when $\Ma _\text{lattice}$ and
also $E_\phi$ ($\phi = u$, $T$, and $\rho$) are very small is probably
caused by the boundary condition treatment.
\par

\subsection{Natural convection in a square cavity}
To further validate the present LB model for coupled thermo-hydrodynamic
flows, the natural convection in a square cavity with a large temperature
difference is simulated in this section. All the four walls of the cavity
are at rest, among which the left (heating) and right (cooling) walls keep
at the temperature $T_h$ and $T_c$ ($T_h > T_c$), respectively, and the
horizontal walls keep adiabatic. The temperature difference between the
heating and cooling walls is quantified by a dimensionless parameter
$\varepsilon = (T_h - T_c) / (2T_0)$, where the reference temperature $T_0 =
(T_h + T_c) /2$. The ideal-gas EOS $p_\EOS^{} = \rho R_g T$ is adopted here,
and thus $C_p - C_v = R_g$. The specific heat ratio $\gamma = C_p / C_v$ and
the Prandtl number $\Prt = C_p \mu / \lambda$ are assumed to be constant.
The dependence of dynamic viscosity on temperature is described by
Sutherland's law as follows \cite{Vierendeels2003}
\begin{equation}
\dfrac{\mu}{\mu^\ast} = \left( \dfrac{T}{T^\ast} \right) ^{3/2}
\dfrac{T^\ast + S}{T + S},
\end{equation}
where $T ^\ast = 273 \text{K}$, $S = 110.5 \text{K}$, and $\mu ^\ast$ is the
dynamic viscosity at $T ^\ast$. As a key dimensionless parameter associated
with natural convection, the Rayleigh number is defined as
\begin{equation}
\Ra = \Prt \dfrac{ |\mathbf{g}| \rho_0^2 (T_h - T_c) L^3}{T_0 \mu_0^2} ,
\end{equation}
where $\mathbf{g}$ is the gravity acceleration, $L$ is the side length of
the square cavity, and $\mu_0$ is the reference dynamic viscosity at $T_0$.
Initially, the ideal gas in the cavity stays still with temperature $T_0$
and density $\rhor$, and then the temperature of the left and right walls
are abruptly changed to $T_h$ and $T_c$, respectively. In the simulations,
the lattice sound speed is set as $c_s = \sqrt{R_g T_0}$, and the basic
parameters are chosen as $|\mathbf{g}| = 9.81 \text{m}/\text{s}^2$, $R_g =
287 \text{J} / (\text{kg} \cdot \text{K})$, $T_0 = 600 \text{K}$, and $\rhor
= p_0^{} / (R_g T_0)$ with $p_0^{} = 101325 \text{Pa}$. The Rayleigh number
$\Ra$ varies from $10^3$ up to $10^8$, while the remaining dimensionless
parameters are fixed at $\varepsilon = 0.6$, $\gamma = 1.4$, and $\Prt =
0.71$. The grid sizes $N_x \times N_y$ and the viscosity ratio $\varsigma /
\nu$ adopted for different $\Ra$ are listed in Table \ref{table.NxNy}, where
$\varsigma / \nu$ is set to $2$ and $4$ for $\Ra = 10^7$ and $10^8$,
respectively, to enhance the numerical stability and it is simply set to $1$
for $\Ra \leq 10^6$. As to the velocity and temperature boundary conditions
on all the four walls, they are realized by the present boundary condition
treatment with first-order extrapolation.
\par

\begin{table}[htbp]
  \centering
  \caption{Grid sizes $N_x \times N_y$ and viscosity ratio $\varsigma / \nu$
  for different Rayleigh numbers $\Ra$.}
  \label{table.NxNy}
  \begin{tabular}{ccc ccc c}
    \hline
    $\Ra$ & $10^3$ & $10^4$ & $10^5$ & $10^6$ & $10^7$ & $10^8$ \\
    \hline
    $N_x \times N_y$ & $128 \times 128$ & $192 \times 192$ &
    $256 \times 256$ & $512 \times 512$ & $1024 \times 1024$ &
    $4096 \times 4096$ \\
    $\varsigma / \nu$ & $1$ & $1$ & $1$ & $1$ & $2$ & $4$ \\
    \hline
  \end{tabular}
\end{table}

\begin{figure}[htbp]
  \centering
  \subfigure{\includegraphics[scale=1,draft=\figdraft]{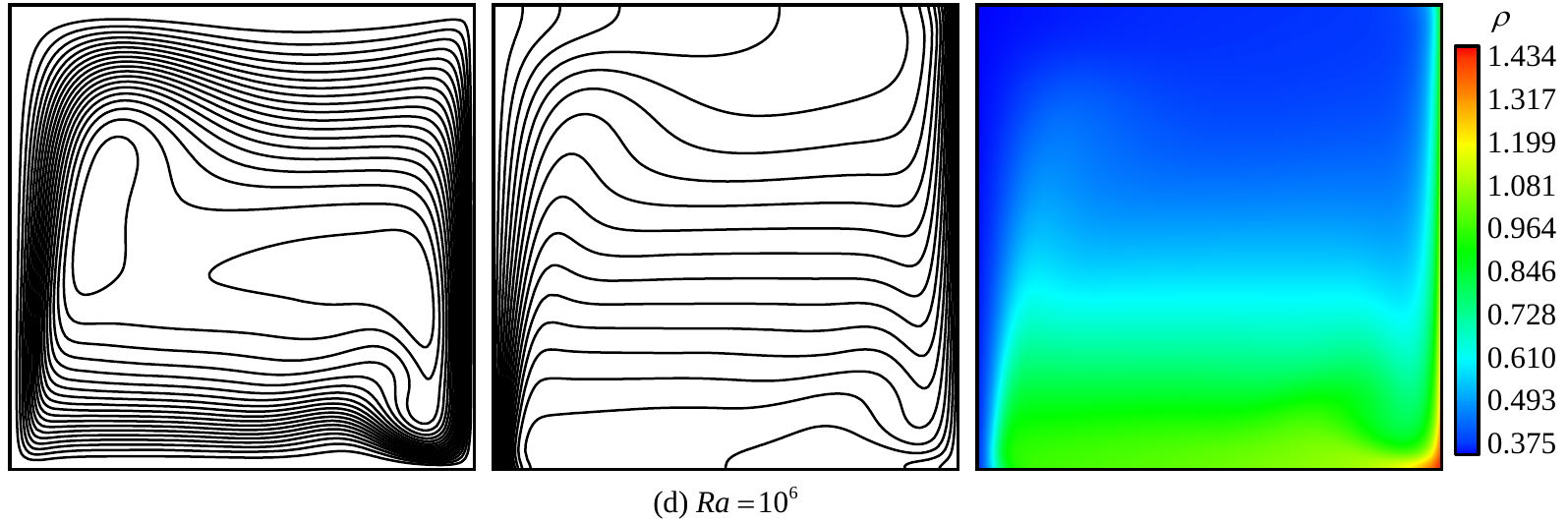}}
  \subfigure{\includegraphics[scale=1,draft=\figdraft]{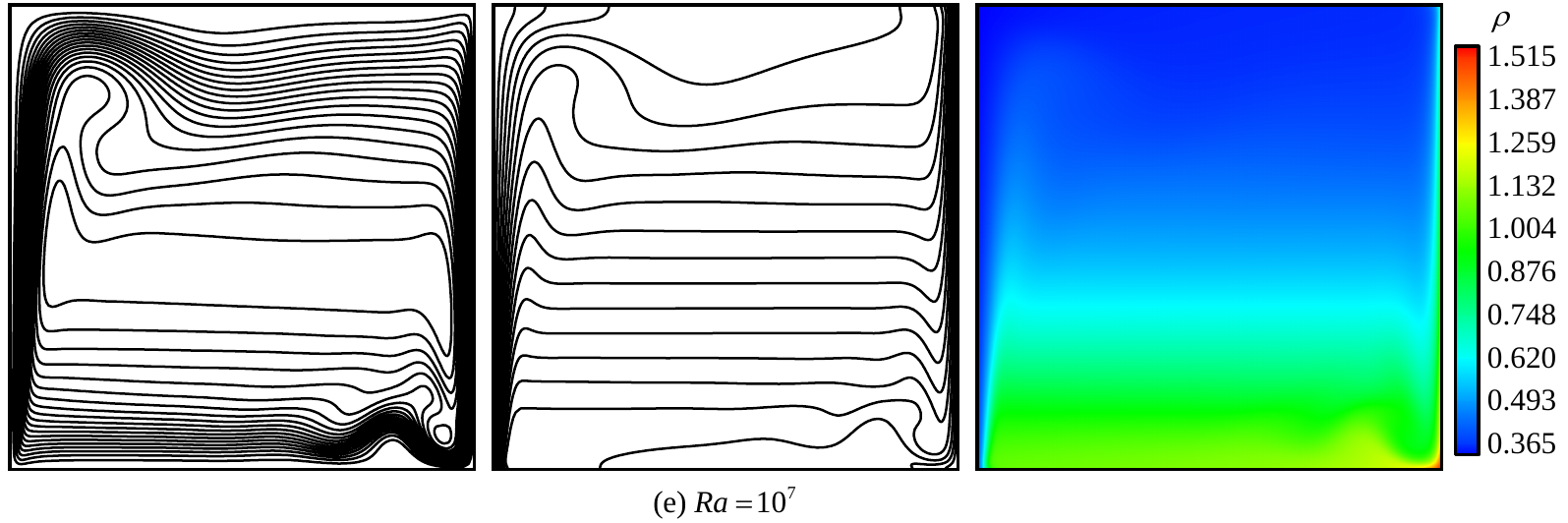}}
  \subfigure{\includegraphics[scale=1,draft=\figdraft]{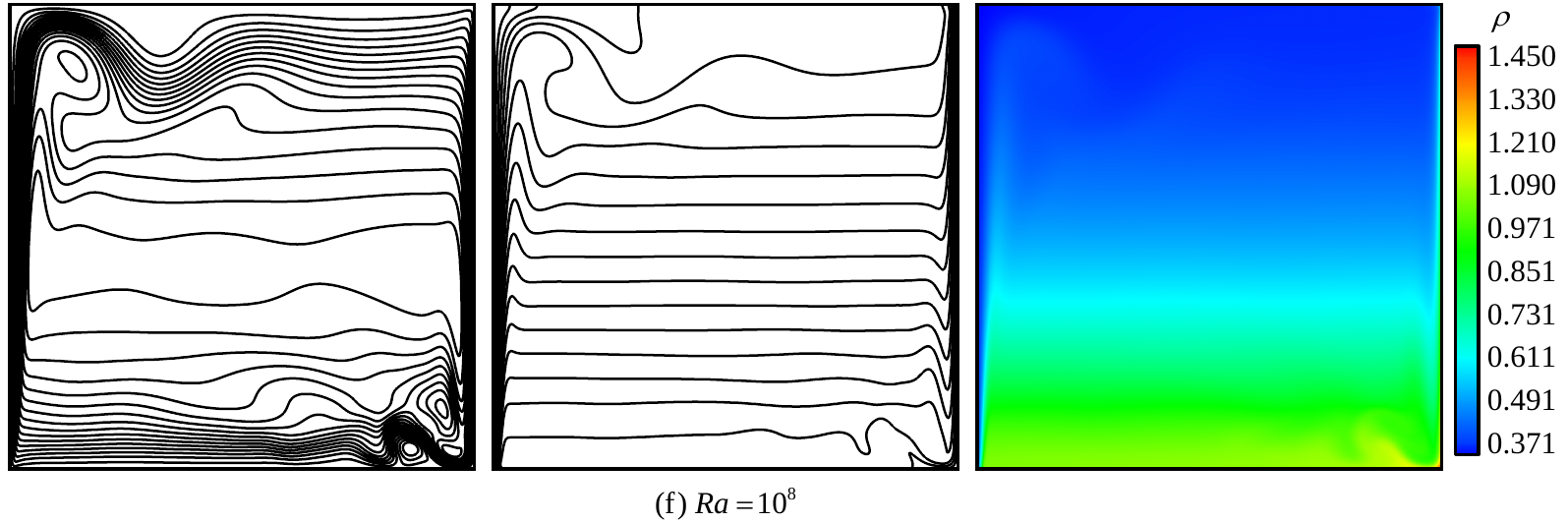}}
  \caption{Streamlines (left), isotherms (middle), and density field (right)
  for natural convection in a square cavity with a large temperature
  difference when $\Ra = 10^3$, $10^4$, $10^5$, $10^6$, $10^7$, and $10^8$.}
  \label{fig.Convection.fields}
\end{figure}

Fig.\ \ref{fig.Convection.fields} shows the streamlines, isotherms, and
density field for the natural convection when $\Ra$ varies from $10^3$ to
$10^8$. It can be seen from Fig.\ \ref{fig.Convection.fields} that a single
vortex with its center closer to the cooling wall appears in the cavity for
$\Ra = 10^3$. As $\Ra$ increases, the vortex is stretched by the natural
convection and breaks up into two vortices when $\Ra = 10^5$. As $\Ra$
further increases, the two vortices move closer to the heating and cooling
walls, respectively, and some small vortices are induced around the center
and in the lower-right and upper-left corners of the cavity when $\Ra =
10^7$. Meanwhile, a counter-rotating vortex also appears in the lower-right
corner and very close to the lower wall for $\Ra = 10^7$. When $\Ra$ reaches
$10^8$, the natural convection becomes unsteady, and many small vortices,
including some counter-rotating ones, are induced by the strong convection.
As to the heat transfer characteristics, it can be seen from Fig.\
\ref{fig.Convection.fields} that the isotherms are almost parallel to the
vertical walls when $\Ra = 10^3$, implying that the heat transfer is
dominated by conduction. As $\Ra$ increases, the isotherms around the cavity
center progressively incline and become parallel to the horizontal walls,
implying that the dominant mechanism for heat transfer changes from
conduction to convection. When $\Ra$ reaches $10^8$, the isotherms spread
along the heating and cooling walls in a very thin layer and become
horizontal almost in the entire cavity. All these observed streamline
patterns and isotherm characteristics are in good agreement with the
previous numerical results \cite{Li2012, Fei2018, Feng2018, Safari2018,
Vierendeels2003}, which are all obtained by the LB method except for the
benchmark solutions reported in Ref.\ \cite{Vierendeels2003}. Note that the
maximum $\Ra$ reported in Refs.\ \cite{Li2012} and \cite{Fei2018} are $10^5$
and $10^6$, respectively, and the maximum $\Ra$ reported in Refs.\
\cite{Feng2018, Safari2018, Vierendeels2003} are $10^7$. On the basis of the
present simulations, it is interesting to find that the natural convection
in a square cavity with a large temperature difference ($\varepsilon = 0.6$)
becomes unsteady when $\Ra = 10^8$, while the corresponding natural
convection with a small temperature difference (i.e., the Boussinesq
approximation is valid) keeps steady when $\Ra = 10^8$ and becomes unsteady
when $\Ra > 1.9 \times 10^8$ \cite{Paolucci1989, Mayne2001, Huang2014.mb}.
From Fig.\ \ref{fig.Convection.fields}, we can also see that the density
significantly varies over space, particularly in the vicinity of the cooling
wall, with its minimum and maximum values smaller and larger than $0.400
\text{kg} / \text{m}^3$ and $1.300 \text{kg} / \text{m}^3$, respectively.
Obviously, the Boussinesq approximation cannot be adopted here. In addition,
the density contours are similar to the isotherms to some extent, which
conforms to the low Mach number condition \cite{Paillere2000}. In fact, the
maximum Mach number is rather small for the natural convection simulated
here \cite{Vierendeels2003}.
\par

\begin{figure}[htbp]
  \centering
  \subfigure{\includegraphics[scale=1,draft=\figdraft]{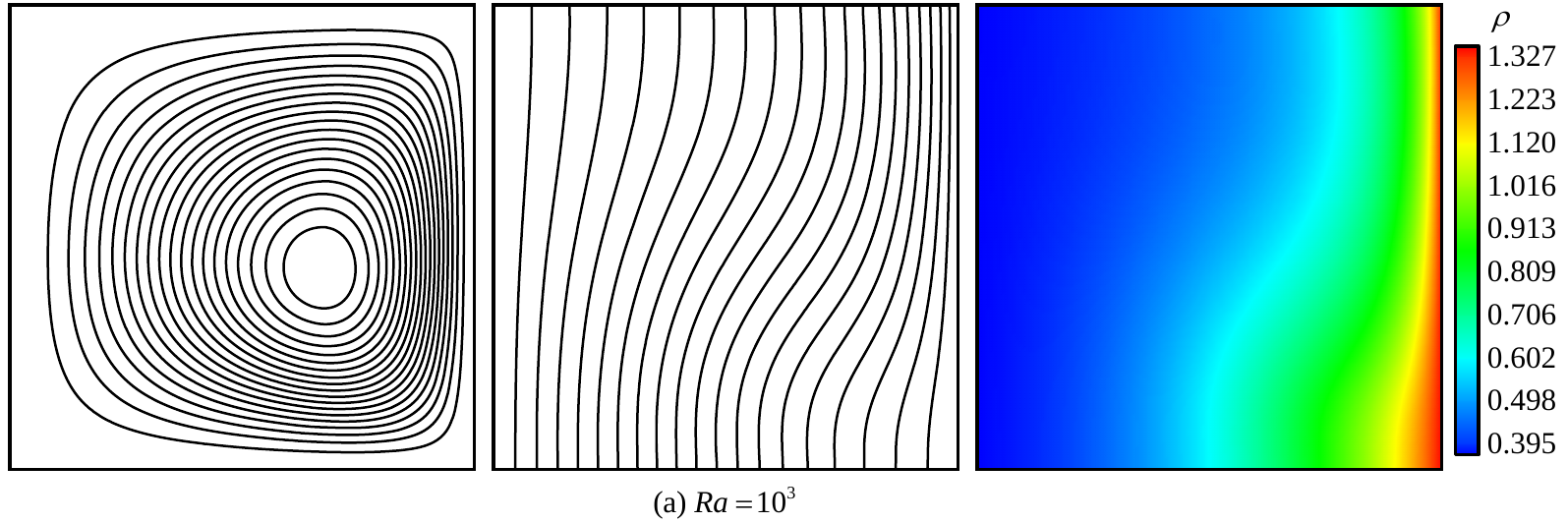}}
  \subfigure{\includegraphics[scale=1,draft=\figdraft]{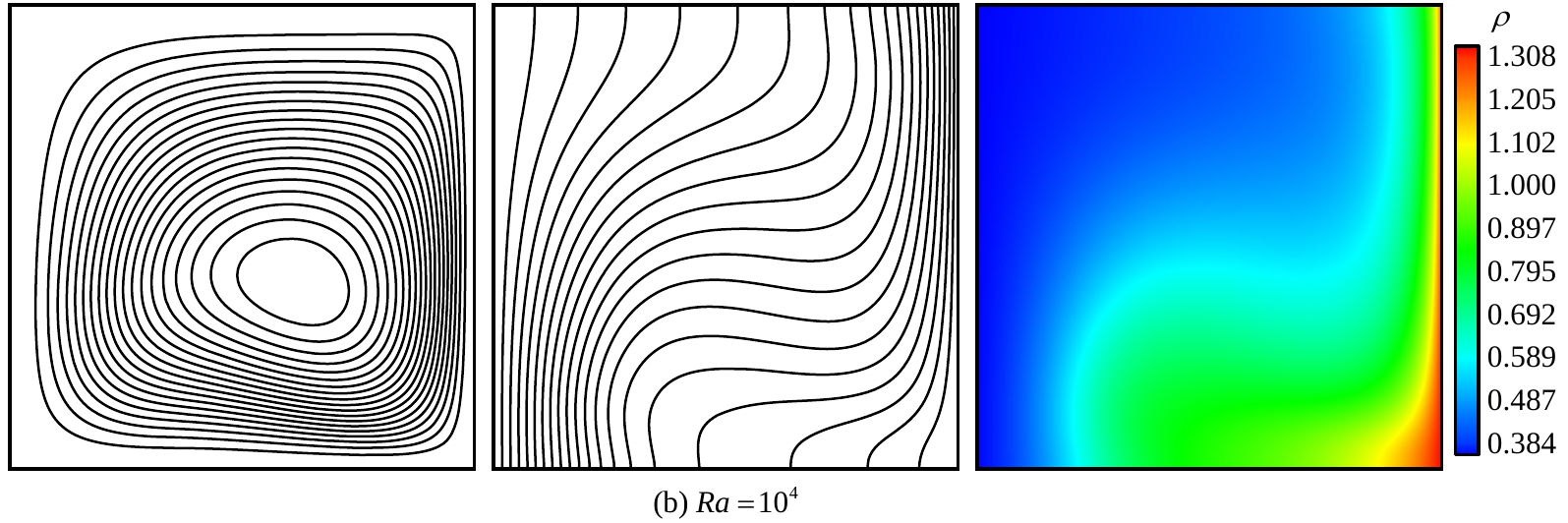}}
  \subfigure{\includegraphics[scale=1,draft=\figdraft]{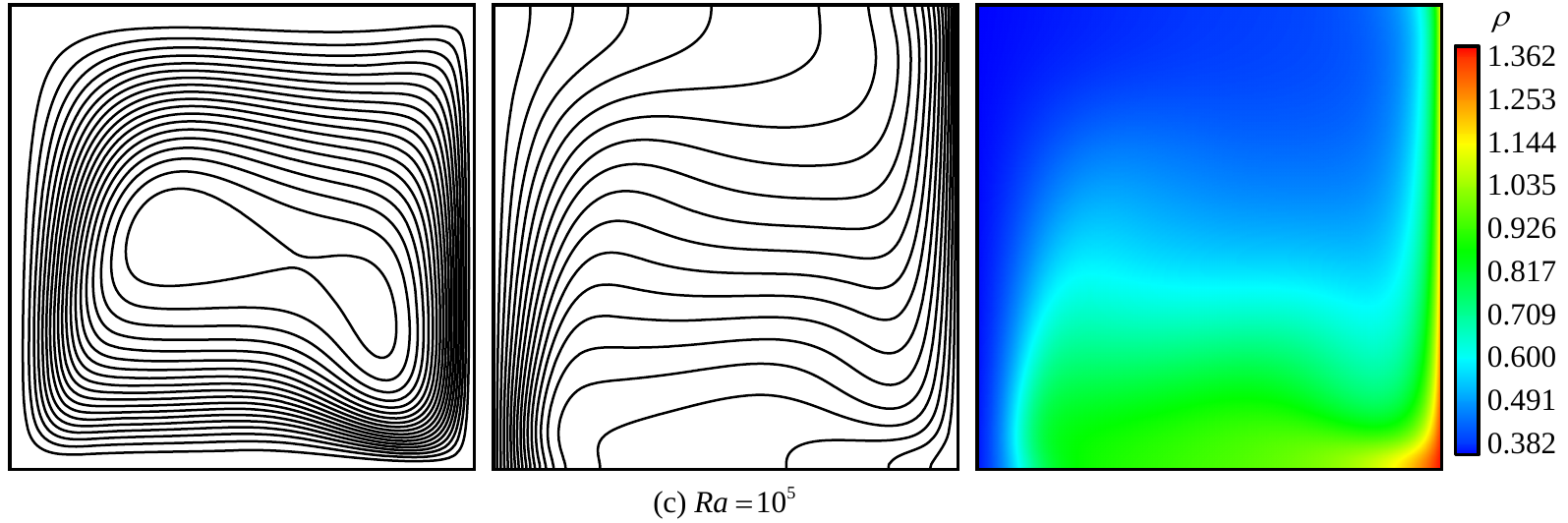}}
\end{figure}

\begin{figure}[htbp]
  \centering
  \includegraphics[scale=1,draft=\figdraft]{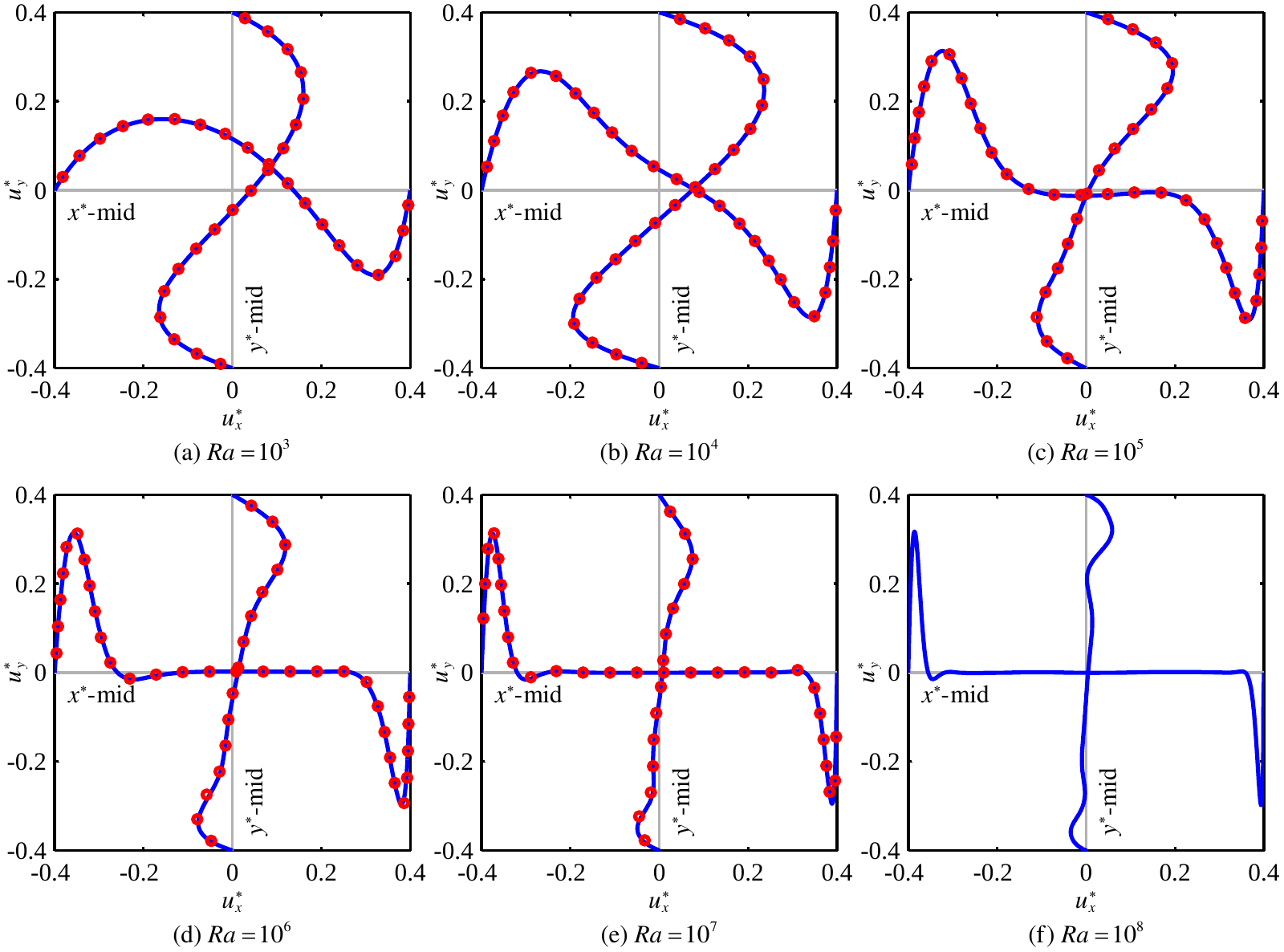}
  \caption{Profiles of the horizontal velocity ($u_x^\ast$) along vertical
  midplane ($y^\ast \text{-mid}$) and the vertical velocity ($u_y^\ast$) along
  horizontal midplane ($x^\ast \text{-mid}$) for natural convection in a
  square cavity with a large temperature difference when $\Ra = 10^3$, $10^4$,
  $10^5$, $10^6$, $10^7$, and $10^8$. The solid lines are the present results
  and the symbols are the benchmark solutions in
  Ref.\ \cite{Vierendeels2003}.}
  \label{fig.Convection.profile}
\end{figure}

To further validate the present results, the profiles of the horizontal
velocity along the vertical midplane and the vertical velocity along the
horizontal midplane are plotted in Fig.\ \ref{fig.Convection.profile} and
compared with the benchmark solutions obtained by Vierendeels et al.\
\cite{Vierendeels2003} using the finite difference (FD) method. Here, the
velocity and coordinate are normalized by the reference velocity
$U_\text{ref} = \sqrt{\Ra} \mspace{1.5mu} \mu_0 \big/ (\rhor L)$ and side
length $L$, respectively, i.e., $\mathbf{u} ^\ast = \mathbf{u} /
U_\text{ref}$ and $\mathbf{x} ^\ast = \mathbf{x} /L$. Note that the
convection becomes unsteady when $\Ra = 10^8$, and thus Fig.\
\ref{fig.Convection.profile}(f) shows the instantaneous profiles at some
time point. Excellent agreement between the present results and the
benchmark solutions can be observed. From Fig.\
\ref{fig.Convection.profile}, we can also see that the velocity profiles are
asymmetric with respect to the cavity center, which is caused by the
invalidation of the Boussinesq approximation. For quantitative comparison,
the average Nusselt number along the heating wall, the average pressure in
the cavity, and the maximum horizontal (vertical) velocity and its position
along the vertical (horizontal) midplane are computed and listed in Table
\ref{Table.Nuave}. Here, the average Nusselt number and pressure are defined
as
\begin{subequations}
\begin{equation}
\Nu_\text{ave} = \dfrac{1}{ \lambda_0 (T_h - T_c) } \int\limits_0^L
J_x (0, y) \text{d}y,
\end{equation}
\begin{equation}
p_\text{ave}^\ast = \dfrac{1}{L^2} \int\limits_0^L \int\limits_0^L
\dfrac{p_\EOS^{} (x, y)}{p_0^{}} \text{d}x\text{d}y ,
\end{equation}
\end{subequations}
where $J_x (x,y)$ is the local heat flux in $x \text{-direction}$,
$\lambda_0^{}$ is the thermal conductivity at $T_0$, and the pressure is
normalized by $p_0^{}$. As seen in Table \ref{Table.Nuave}, the present
results agree well with the previous numerical results, which further
demonstrates the applicability and accuracy of the present LB model for
coupled thermo-hydrodynamic flows.
\par

\begin{table}[htbp]
  \centering
  \caption{Comparisons of the average Nusselt number along the heating wall
  ($\Nu_\text{ave}$), the average pressure in the cavity
  ($p_\text{ave}^\ast$), the maximum horizontal velocity
  ($|u_x^\ast|_\text{max}$) and its position ($y_\text{max}^\ast$)
  along the vertical midplane, and the maximum vertical velocity
  ($|u_y^\ast|_\text{max}$) and its position ($x_\text{max}^\ast$)
  along the horizontal midplane between the present and previous results.}
  \label{Table.Nuave}
  \newlength{\hj}\setlength{\hj}{1.5ex}
  \setlength{\tabcolsep}{13.28pt} %
  \begin{tabular}{ccc ccc cc}
    \hline
    $\Ra$   & Method & $\quad \Nu_\text{ave}$   & $\quad p_\text{ave}^\ast$
               & $|u_x^\ast| _\text{max}$ & $y_\text{max}^\ast$
               & $|u_y^\ast| _\text{max}$ &  $x_\text{max}^\ast$    \\
    \hline
    $10^3$ & Present                                     & $1.1063$
           & $0.93443$ & $0.1653$ & $0.1641$ & $0.1911$  & $0.8984$ \\
  	~      & FD method \cite{Vierendeels2003}            & $1.1077$
           & $0.93805$ & $0.1649$ & $0.1618$ & $0.1926$  & $0.9036$ \\
    ~      & LB method \cite{Fei2018}                    & $1.106$
           & --        & $0.1639$ & $0.1624$ & $0.1925$  & $0.9063$ \\
    ~      & LB method \cite{Li2012}                     & $1.111$
           & --        & $0.1660$ & $0.1600$ & $0.1973$  & $0.9100$
                                                         \vspace{\hj} \\
    $10^4$ & Present                                     & $2.2123$
           & $0.91144$ & $0.2360$ & $0.7813$ & $0.2857$  & $0.9271$ \\
    ~      & FD method \cite{Vierendeels2003}            & $2.218$
           & $0.91463$ & $0.2363$ & $0.7821$ & $0.2863$  & $0.9270$ \\
    ~      & LB method \cite{Fei2018}                    & $2.224$
           & --        & $0.2372$ & $0.7813$ & $0.2859$  & $0.9312$ \\
    ~      & LB method \cite{Li2012}                     & $2.217$
           & --        & $0.2364$ & $0.7800$ & $0.2874$  & $0.9267$
                                                         \vspace{\hj} \\
    $10^5$ & Present                                     & $4.4836$
           & $0.91719$ & $0.1950$ & $0.8359$ & $0.3130$  & $0.0977$ \\
    ~      & FD method \cite{Vierendeels2003}            & $4.480$
           & $0.92196$ & $0.1946$ & $0.8364$ & $0.3166$  & $0.0948$ \\
    ~      & LB method \cite{Fei2018}                    & $4.512$
           & --        & $0.1951$ & $0.8344$ & $0.3176$  & $0.0938$ \\
    ~      & LB method \cite{Li2012}                     & $4.454$
           & --        & $0.1959$ & $0.8360$ & $0.3165$  & $0.0960$
                                                         \vspace{\hj} \\
    $10^6$ & Present                                     & $8.7406$
           & $0.91820$ & $0.1193$ & $0.8516$ & $0.3141$  & $0.0547$ \\
    ~      & FD method \cite{Vierendeels2003}            & $8.687$
           & $0.92449$ & $0.1193$ & $0.8541$ & $0.3203$  & $0.0537$ \\
    ~      & LB method \cite{Fei2018}                    & $8.691$
           & --        & $0.1202$ & $0.8551$ & $0.3159$  & $0.0540$
                                                         \vspace{\hj} \\
    $10^7$ & Present                                     & $16.4373$
           & $0.91425$ & $0.0745$ & $0.8262$ & $0.3124$  & $0.0313$ \\
    ~      & FD method \cite{Vierendeels2003}            & $16.240$
           & $0.92263$ & $0.0749$ & $0.8260$ & $0.3229$  & $0.0305$
                                                         \vspace{\hj} \\
    $10^8$ & Present                                     & $29.9435$
           & $0.91609$ & $0.0586$ & $0.9004$ & $0.3169$  & $0.0176$ \\
    \hline
  \end{tabular}
\end{table}

\section{Conclusions} \label{sec.conclusion}
A novel LB model for coupled thermo-hydrodynamic flows is developed in the
framework of the DDF approach. The velocity field is solved by the recently
developed LB equation for density DF, by which the recovered EOS can be
self-tuned via a built-in variable, implying that various EOSs can be
adopted in real applications. With the energy conservation equation properly
reformulated, a novel LB equation for total energy DF is directly developed
at the discrete level to solve the temperature field. The viscous
dissipation is recovered along with the conduction term by introducing a
density-DF-related term into this LB equation, while the compression work is
recovered along with the convection term by devising the equilibrium moment
function for total energy DF. The work done by force is absorbed into the
source term and then correctly incorporated into the LB equation via the
discrete source term. Moreover, by modifying the collision matrix, the
targeted energy conservation equation can be recovered without deviation
term. The development of the present LB model, with double MRT collision
schemes employed, is based on the standard lattice, and both the Prandtl
number and specific heat ratio can be arbitrarily adjusted. On the basis of
judiciously decomposing DF into its equilibrium, force (source), and
nonequilibrium parts, boundary condition treatment is further proposed for
simulating coupled thermo-hydrodynamic flows, which can ensure the local
conservation of mass, momentum, and energy at the boundary node. The
applicability and accuracy of the present LB model with self-tuning EOS are
first validated by simulating thermal Poiseuille and Couette flows with the
decoupling, ideal-gas, and Carnahan-Starling EOSs. Then, the present LB
model is successfully applied to the simulation of natural convection in a
square cavity with a large temperature difference for the Rayleigh number
ranging from $10^3$ up to $10^8$, and the obtained results agree very well
with the previous benchmark solutions.

\section*{Acknowledgements}
R.H.\ acknowledges the support by the Alexander von Humboldt Foundation,
Germany. This work was also supported by the National Natural Science
Foundation of China through Grant No.\ 51536005.

\appendix

\section{Chapman-Enskog analysis} \label{app.CE}
The detailed Chapman-Enskog analysis of the LB equation for density DF
(i.e., Eq.\ (\ref{eq.LBE.u})) can be found in our previous work
\cite{Huang2018.eos}. Here, the Chapman-Enskog analysis of the LB equation
for total energy DF (i.e., Eq.\ (\ref{eq.LBE.T})) is carried out to recover
the corresponding macroscopic conservation equation. For this purpose,
performing the Taylor series expansion of $g_i (\mathbf{x} + \mathbf{e}_i
\delta_t, t+ \delta_t)$ centered at $(\mathbf{x}, t)$ in Eq.\
(\ref{eq.LBE.T.s}), and then transforming the result into moment space and
combining it with Eq.\ (\ref{eq.LBE.T.c}), we have
\begin{equation}\label{eq.LBE.Taylor}
(\mathbf{I} \partial_t + \mathbf{D}) \mathbf{n} + \dfrac{\delta_t}{2}
(\mathbf{I} \partial_t + \mathbf{D})^2 \mathbf{n} -
\mathbf{Q}_m + O(\delta_t^2) = - \dfrac{\mathbf{L}}{\delta_t} \left(
\mathbf{n} - \mathbf{n}^\eq + \dfrac{\delta_t}{2} \mathbf{Q}_m \right) +
\dfrac{c^2 \mathbf{Y}}{\delta_t} \left(
\dfrac{\mathbf{m} + \bbm}{2} - \mathbf{m}^\eq \right),
\end{equation}
where $\mathbf{D} = \mathbf{M} [\text{diag} (\mathbf{e}_i \cdot \nabla)]
\mathbf{M} ^{-1}$. With the following Chapman-Enskog expansions
\cite{Chapman1970}
\begin{subequations}
\begin{equation}
\partial_t = \sum\limits_{n=1}^{+\infty} \varepsilon^n \partial_{tn},
\quad
\nabla = \varepsilon^1 \nabla_1, \quad
q = \varepsilon^1 q^{(1)}, \quad
\mathbf{F} = \varepsilon^1 \mathbf{F}^{(1)} ,
\end{equation}
\begin{equation}
\mathbf{n}= \sum\limits_{n=0}^{+\infty} \varepsilon^n \mathbf{n}^{(n)},
\quad
\mathbf{m}= \sum\limits_{n=0}^{+\infty} \varepsilon^n \mathbf{m}^{(n)},
\quad
\bbm = \sum\limits_{n=0}^{+\infty} \varepsilon^n \bbm^{(n)},
\end{equation}
\end{subequations}
we have $\mathbf{D} = \varepsilon^1 \mathbf{D}_1$, $q_e = \varepsilon^1
q_e^{(1)}$, and $\mathbf{Q}_m = \varepsilon^1 \mathbf{Q}_m^{(1)}$, where
$\varepsilon$ is the small expansion parameter. Substituting these
expansions into Eq.\ (\ref{eq.LBE.Taylor}), the $\varepsilon^0 \text{-}$,
$\varepsilon^1 \text{-}$, and $\varepsilon^2 \text{-order}$ equations can
then be obtained as
\begin{subequations}
\begin{equation}\label{eq.CE.order0}
\varepsilon^0 :
- \dfrac{ \mathbf{L} }{\delta_t} (\mathbf{n}^{(0)} - \mathbf{n} ^\eq) +
\dfrac{c^2 \mathbf{Y}}{\delta_t} \left(
\dfrac{ \mathbf{m}^{(0)} + \bbm^{(0)}}{2} - \mathbf{m} ^\eq
\right) = \mathbf{0},
\end{equation}
\begin{equation}\label{eq.CE.order1}
\varepsilon^1 :
( \mathbf{I} \partial_{t1} + \mathbf{D}_1 ) \mathbf{n} ^{(0)} -
\mathbf{Q}_m^{(1)} = -\dfrac{\mathbf{L}}{\delta_t} \mathbf{H} ^{(1)} +
\dfrac{c^2 \mathbf{Y}}{\delta_t}
\dfrac{ \mathbf{m}^{(1)} + \bbm^{(1)} }{2} ,
\end{equation}
\begin{equation}\label{eq.CE.order2}
\varepsilon^2 :
\partial_{t2} \mathbf{n} ^{(0)} + (\mathbf{I} \partial_{t1} +
\mathbf{D}_1)
\left[ \left( \mathbf{I} - \dfrac{\mathbf{L}}{2} \right) \mathbf{H}^{(1)}
+
\dfrac{c^2 \mathbf{Y}}{2} \dfrac{ \mathbf{m}^{(1)} + \bbm^{(1)}}{2}
\right] = -\dfrac{\mathbf{L}}{\delta_t} \mathbf{H} ^{(2)} +
\dfrac{c^2 \mathbf{Y}}{\delta_t}
\dfrac{ \mathbf{m}^{(2)} + \bbm^{(2)} }{2} ,
\end{equation}
\end{subequations}
where $\mathbf{H}^{(1)} = \mathbf{n} ^{(1)} + \delta_t \mathbf{Q}_m^{(1)}
\big/ 2$ is introduced to simplify the descriptions.
\par

From the $\varepsilon^0 \text{-order}$ equation (i.e., Eq.\
(\ref{eq.CE.order0})) and considering $\mathbf{m} ^{(0)} = \bbm ^{(0)} =
\mathbf{m} ^\eq$ (see Eq.\ (\ref{eq.m.order1})), we have
\begin{equation}
\varepsilon^0 : \mathbf{n} ^{(0)} = \mathbf{n} ^\eq ,
\end{equation}
which indicates that the $\varepsilon^n \text{-order}$ ($n \geq 1$) terms of
the conserved moment $n_0$ satisfy
\begin{equation}\label{eq.CE.order0.n0}
H_0^{(1)} = n_0^{(1)} + \dfrac{\delta_t}{2} Q_{m,0}^{(1)} = 0, \quad
n_0^{(n)} = 0 \; (\forall n \geq 2) .
\end{equation}
The $\varepsilon^1 \text{-order}$ equation for $n_0$, extracted from Eq.\
(\ref{eq.CE.order1}), is given as
\begin{equation}
\varepsilon^1 : \partial_{t1} n_0^{(0)} + c \nabla_1 \cdot
\begin{bmatrix}
n_3^{(0)}  \\   n_5^{(0)} \\
\end{bmatrix}
- Q_{m,0}^{(1)} = - \dfrac{\sigma_0^{}}{\delta_t} H_0^{(1)},
\end{equation}
which can be simplified as follows
\begin{equation}\label{eq.CE.order1.rhoE}
\varepsilon^1 :
\partial_{t1} (\rho E) + \nabla_1 \cdot (\rho H \mathbf{u}) - q_e^{(1)} = 0 .
\end{equation}
Similarly, the $\varepsilon^2 \text{-order}$ equation for $n_0$, extracted
from Eq.\ (\ref{eq.CE.order2}), is given as
\begin{equation}\label{eq.CE.order2.n0}
\varepsilon^2 :
\partial_{t2} n_0^{(0)} + \partial_{t1} \left[ \left( 1 -
\dfrac{\sigma_0^{}}{2} \right) H_0^{(1)} \right] + c \nabla_1 \cdot
\left\{ \left( 1 - \dfrac{\sigma_j^{}}{2} \right)
\begin{bmatrix}
H_3^{(1)} + \tfrac{\sigma_q}{2} H_4^{(1)} \\
H_5^{(1)} + \tfrac{\sigma_q}{2} H_6^{(1)}
\end{bmatrix}
+ \dfrac{c^2}{2}
\begin{bmatrix}
Y_{3\alpha}    \\    Y_{5\alpha}
\end{bmatrix}
\dfrac{ m_\alpha^{(1)} + \bar{m}_\alpha^{(1)} }{2}
\right\}  = - \dfrac{\sigma_0^{}}{\delta_t} n_0^{(2)} ,
\end{equation}
where the repeated index implies summation from $0$ to $8$. With Eq.\
(\ref{eq.CE.order0.n0}), Eq.\ (\ref{eq.CE.order2.n0}) can be simplified as
follows
\begin{equation}\label{eq.CE.order2.rhoE}
\varepsilon^2 :
\partial_{t2} (\rho E) = \nabla_1 \cdot \mathbf{J} ^{(1)},
\end{equation}
where $\mathbf{J} ^{(1)}$ is the energy flux expressed as
\begin{equation}\label{eq.CE.order1.J}
\mathbf{J} ^{(1)} = -c \left( 1 - \dfrac{\sigma_j^{}}{2} \right)
\begin{bmatrix}
H_3^{(1)} + \tfrac{\sigma_q}{2} H_4^{(1)} \\
H_5^{(1)} + \tfrac{\sigma_q}{2} H_6^{(1)}
\end{bmatrix}
- \dfrac{c^3}{2}
\begin{bmatrix}
Y_{3\alpha}    \\    Y_{5\alpha}
\end{bmatrix}
\dfrac{ m_\alpha^{(1)} + \bar{m}_\alpha^{(1)} }{2} ,
\end{equation}
in which the first and second terms will account for the heat conduction and
viscous dissipation, respectively.
\par

To simplify the heat conduction term in Eq.\ (\ref{eq.CE.order1.J}), we add
the $\varepsilon^1 \text{-order}$ equation for $n_4$ to the $\varepsilon^1
\text{-order}$ equation for $n_3$ and the $\varepsilon^1 \text{-order}$
equation for $n_6$ to the $\varepsilon^1 \text{-order}$ equation for $n_5$,
and then combine the results together and finally have
\begin{equation}\label{eq.CE.order1.J.T}
\begin{split}
-\dfrac{\sigma_j^{}}{\delta_t}
\begin{bmatrix}
H_3^{(1)} + \tfrac{\sigma_q}{2} H_4^{(1)} \\
H_5^{(1)} + \tfrac{\sigma_q}{2} H_6^{(1)}
\end{bmatrix}
= \begin{bmatrix}
\partial_{t1} \left( n_3^{(0)} + n_4^{(0)} \right) + c \partial_{x1}
\left( \tfrac23 n_0^{(0)}
+ \tfrac12 n_1^{(0)} + \tfrac13 n_2^{(0)} - \tfrac12 n_7^{(0)} \right) +
2c \partial_{y1} n_8^{(0)} \\
\partial_{t1} \left( n_5^{(0)} + n_6^{(0)} \right) +
2c \partial_{x1} n_8^{(0)} +
c \partial _{y1} \left( \tfrac23 n_0^{(0)} + \tfrac12 n_1^{(0)}
+ \tfrac13 n_2^{(0)} + \tfrac12 n_7^{(0)} \right)
\end{bmatrix}  &  \\
-
\begin{bmatrix}
Q_{m,3}^{(1)} + Q_{m,4}^{(1)} \\
Q_{m,5}^{(1)} + Q_{m,6}^{(1)}
\end{bmatrix}
- \dfrac{c^2}{\delta_t}
\begin{bmatrix}
Y_{3\alpha} + Y_{4\alpha} \\
Y_{5\alpha} + Y_{6\alpha}
\end{bmatrix}
\dfrac{ m_\alpha^{(1)} + \bar{m}_\alpha^{(1)}}{2}  &.
\end{split}
\end{equation}
Considering $Y_{3\alpha} + Y_{4\alpha} =0$ and $Y_{5\alpha} + Y_{6\alpha}
=0$, Eq.\ (\ref{eq.CE.order1.J.T}) can be simplified as
\begin{equation}\label{eq.CE.order1.J.T.final}
-\dfrac{\sigma_j^{}}{\delta_t}
\begin{bmatrix}
H_3^{(1)} + \tfrac{\sigma_q}{2} H_4^{(1)} \\
H_5^{(1)} + \tfrac{\sigma_q}{2} H_6^{(1)}
\end{bmatrix} = \left( \dfrac23 + \dfrac{\gamma_1^{}}{2}
+ \dfrac{\gamma_2^{}}{3} \right) c \nabla_1 (\rhor C_{p,0} T) ,
\end{equation}
where $\rhor C_{p,0}$ and $\nabla_1$ are commutative, implying that the heat
conduction is correctly driven by the temperature gradient. To recover the
viscous dissipation term via Eq.\ (\ref{eq.CE.order1.J}), we can directly
set
\begin{equation}\label{eq.CE.order1.J.Pi}
-\dfrac{c^3}{2}
\begin{bmatrix}
Y_{3\alpha}    \\    Y_{5\alpha}
\end{bmatrix}
\dfrac{ m_\alpha^{(1)} + \bar{m}_\alpha^{(1)}}{2} =
\mathbf{u} \cdot \mathbf{\Pi}^{(1)} ,
\end{equation}
where the viscous stress tensor $\mathbf{\Pi}^{(1)}$ is given by Eq.\
(\ref{eq.Pi}). Consequently, the nonzero elements in $\mathbf{Y}$ can be
completely and uniquely determined (see Eq.\ (\ref{eq.Y.nonzero})). On the
basis of Eqs.\ (\ref{eq.CE.order1.J.T.final}) and (\ref{eq.CE.order1.J.Pi}),
$\mathbf{J} ^{(1)}$ can be written as
\begin{equation}\label{eq.CE.order1.J.final}
\mathbf{J} ^{(1)} = \left( \dfrac23 + \dfrac{\gamma_1^{}}{2} +
\dfrac{\gamma_2^{}}{3} \right) \rhor C_{p,0} c^2 \delta_t \left(
\dfrac{1}{\sigma_j^{}} - \dfrac12 \right) \nabla_1 T +
\mathbf{u} \cdot \mathbf{\Pi}^{(1)} .
\end{equation}
\par

Combining the $\varepsilon^1 \text{-}$ and $\varepsilon^2 \text{-order}$
equations (i.e., Eqs.\ (\ref{eq.CE.order1.rhoE}) and
(\ref{eq.CE.order2.rhoE})), as well as considering Eq.\
(\ref{eq.CE.order1.J.final}), the following macroscopic conservation
equation can be recovered
\begin{equation}\label{eq.CE.ECE}
\partial_t (\rho E) + \nabla \cdot (\rho H \mathbf{u}) = \nabla \cdot
\left[
\left( \dfrac23 + \dfrac{\gamma_1^{}}{2} + \dfrac{\gamma_2^{}}{3} \right)
\rhor C_{p,0} c^2 \delta_t \left( \dfrac{1}{\sigma_j^{}} - \dfrac12
\right)
\nabla T + \mathbf{u} \cdot \mathbf{\Pi} \right] + q_e.
\end{equation}
Therefore, the heat conductivity is $\lambda = ( 2/3 + \gamma_1^{}/2 +
\gamma_2^{}/3 ) \rhor C_{p,0} c^2 \delta_t \big( \sigma_j^{-1} - 0.5 \big)$.
Compared with the targeted energy conservation equation (i.e., Eq.\
(\ref{eq.ECE})), no deviation term exists in Eq.\ (\ref{eq.CE.ECE}) due to
the modification of the collision matrix $\mathbf{L}$.
\par

\section{Inverse matrix} \label{app.Inverse}
The inverse matrix of $\mathbf{I} - \mathbf{S}/2$ is
\begin{equation}
\left( \mathbf{I} - \dfrac{\mathbf{S}}{2} \right) ^{-1} =
\begin{bmatrix}
1 & 0 & 0 & 0 & 0 & 0 & 0 & 0 & 0 \\
0 & 1 & -k s_\varepsilon^{} /2 & 0 & -h \hu_x s_q^{} /2 & 0
                                   & -h \hu_y s_q^{} /2 & 0 & 0 \\
0 & 0 & 1 & 0 & 0 & 0 & 0 & 0 & 0 \\
0 & 0 & 0 & 1 & 0 & 0 & 0 & 0 & 0 \\
0 & 0 & 0 & 0 & 1 & 0 & 0 & 0 & 0 \\
0 & 0 & 0 & 0 & 0 & 1 & 0 & 0 & 0 \\
0 & 0 & 0 & 0 & 0 & 0 & 1 & 0 & 0 \\
0 & 0 & 0 & 0 & -b \hu_x s_q^{} & 0
              &  b \hu_y s_q^{} & 1 & 0 \\
0 & 0 & 0 & 0 & -b \hu_y s_q^{} /2 & 0
              & -b \hu_x s_q^{} /2 & 0 & 1 \\
\end{bmatrix}
\left[ \mathbf{I} - \dfrac{\text{diag}(\mathbf{S})}{2} \right] ^{-1} ,
\end{equation}
where $\text{diag} (\mathbf{S})$ denotes the diagonal part of $\mathbf{S}$.
The inverse matrix of $\mathbf{I} - \mathbf{L}/2$ is
\begin{equation}
\left( \mathbf{I} - \dfrac{\mathbf{L}}{2} \right) ^{-1} =
\begin{bmatrix}
1 & 0 & 0 & 0 & 0 & 0 & 0 & 0 & 0 \\
0 & 1 & 0 & 0 & 0 & 0 & 0 & 0 & 0 \\
0 & 0 & 1 & 0 & 0 & 0 & 0 & 0 & 0 \\
0 & 0 & 0 & 1 & -\sigma_q^{} /2 & 0 & 0 & 0 & 0 \\
0 & 0 & 0 & 0 & 1 & 0 & 0 & 0 & 0 \\
0 & 0 & 0 & 0 & 0 & 1 & -\sigma_q^{} /2 & 0 & 0 \\
0 & 0 & 0 & 0 & 0 & 0 & 1 & 0 & 0 \\
0 & 0 & 0 & 0 & 0 & 0 & 0 & 1 & 0 \\
0 & 0 & 0 & 0 & 0 & 0 & 0 & 0 & 1 \\
\end{bmatrix}
\left[ \mathbf{I} - \dfrac{\text{diag}(\mathbf{L})}{2} \right] ^{-1} ,
\end{equation}
where $\text{diag} (\mathbf{L})$ denotes the diagonal part of $\mathbf{L}$.
\par

\section{Implementation}\label{app.Implementation}
The detailed implementation of the collision process for density DF (i.e.,
Eq.\ (\ref{eq.LBE.u.c})) can be found in our previous work
\cite{Huang2018.eos}. Here, a similar implementation of the collision
process for total energy DF (i.e., Eq.\ (\ref{eq.LBE.T.c})) is given. In
real applications, Eq.\ (\ref{eq.LBE.T.c}) can be executed in the following
sequence
\begin{enumerate}[(1)]
  \item \label{step.1} $ \begin{cases}
  \bar{\mathbf{n}} \leftarrow \mathbf{n} ,\\
  \mathbf{n} \leftarrow \mathbf{n} - \mathbf{n} ^\eq ,\\
  \bar{\mathbf{n}} \leftarrow \bar{\mathbf{n}} - 2 \mathbf{n} ,\\
  \mathbf{n} \leftarrow \mathbf{n} + \delta_t \mathbf{Q}_m /2 ;
  \end{cases} $
  \item \label{step.2} $ \begin{cases}
  n_3 \leftarrow n_3 + \sigma_q^{} n_4 /2 , \\
  n_5 \leftarrow n_5 + \sigma_q^{} n_6 /2 ;
  \end{cases} $
  \item \label{step.3} $ \begin{cases} \mathbf{n} \leftarrow [\mathbf{I} -
      \text{diag}(\mathbf{L})/2]
  \mathbf{n} ,\\
  \bar{\mathbf{n}} \leftarrow \bar{\mathbf{n}} + 2 \mathbf{n} ;
  \end{cases} $
  \item \label{step.4} $ \begin{cases} \bar{n}_3 \leftarrow \bar{n}_3 +
      c^2 Y_{31} \left[ (m_1+\bar{m}_1)/2 - m_1^\eq \right] + c^2 Y_{37}
      \left[ (m_7+\bar{m}_7)/2 - m_7^\eq \right] + c^2 Y_{38} \left[
      (m_8+\bar{m}_8)/2 - m_8^\eq \right] , \\
  \bar{n}_4 \leftarrow \bar{n}_4 + c^2 Y_{41} \left[ (m_1+\bar{m}_1)/2 -
m_1^\eq \right] + c^2 Y_{47} \left[ (m_7+\bar{m}_7)/2 - m_7^\eq \right] +
c^2 Y_{48} \left[
(m_8+\bar{m}_8)/2 - m_8^\eq \right] , \\
      \bar{n}_5 \leftarrow \bar{n}_5 + c^2 Y_{51} \left[ (m_1+\bar{m}_1)/2
- m_1^\eq \right] + c^2 Y_{57} \left[ (m_7+\bar{m}_7)/2 - m_7^\eq \right]
+ c^2 Y_{58} \left[
(m_8+\bar{m}_8)/2 - m_8^\eq \right] , \\
      \bar{n}_6 \leftarrow \bar{n}_6 + c^2 Y_{61} \left[ (m_1+\bar{m}_1)/2
- m_1^\eq \right] + c^2 Y_{67} \left[ (m_7+\bar{m}_7)/2 - m_7^\eq \right]
+ c^2 Y_{68} \left[
(m_8+\bar{m}_8)/2 - m_8^\eq \right] ; \\
  \end{cases} $
\end{enumerate}
where ``$\leftarrow$'' indicates assignment, and steps (\ref{step.2}) and
(\ref{step.4}) correspond to the modification of collision matrix and the
consideration of viscous dissipation, respectively. In step (\ref{step.4}),
$(m_\alpha + \bar{m}_\alpha) /2 - m_\alpha ^\eq$ ($\alpha = 1$, $7$, and
$8$) can be directly obtained from the collision process for density DF.
From the above discussion, it can be seen that the present collision process
is easy to implement with high efficiency although the modified collision
matrix is nondiagonal.
\par

\biboptions{sort&compress}

\end{document}